\documentclass[aps,prd,twocolumn,prd,showpacs,showkeys,preprintnumbers,bibnotes,floatfix,longbibliography,notitlepage,nofootinbib,superscriptaddress,
]{revtex4-2}

\pdfoutput=1
\usepackage{amsmath}
\usepackage{amsfonts}
\usepackage{amssymb}
\usepackage{mathrsfs}
\usepackage{graphicx}
\usepackage{color}
\usepackage{bbding}
\usepackage{tabularx}
\usepackage[usenames,dvipsnames,table]{xcolor}
\usepackage[normalem]{ulem}
\usepackage{braket}
\usepackage{makecell}
\usepackage{slashed}
\usepackage{bm}
\usepackage{booktabs}
\usepackage{xurl}
\usepackage[hidelinks]{hyperref}
\usepackage{multirow}
\usepackage{float}
\usepackage{makecell}
\usepackage{upgreek}
\usepackage[capitalise]{cleveref}

\newcommand{\revAdd}[1]{\textcolor{black}{#1}}

\begin{document}

\title{Novel Light Dark Matter Detection with Quantum Parity Detector Using Qubit Arrays}

\author{Xuegang Li}
\affiliation{Beijing Key Laboratory of Fault-Tolerant Quantum Computing, Beijing Academy of Quantum Information Sciences, Beijing, China}

\author{Yuxiang Liu}
\affiliation{School of Physics and State Key Laboratory of Nuclear Physics and Technology, Peking University, Beijing 100871, China}
\affiliation{Beijing Key Laboratory of Fault-Tolerant Quantum Computing, Beijing Academy of Quantum Information Sciences, Beijing, China}

\author{Jing Shu}
\thanks{Corresponding author.\\ jshu@pku.edu.cn}
\affiliation{School of Physics and State Key Laboratory of Nuclear Physics and Technology, Peking University, Beijing 100871, China}
\affiliation{Center for High Energy Physics, Peking University, Beijing 100871, China}
\affiliation{Beijing Laser Acceleration Innovation Center, Huairou, Beijing, 101400, China}

\author{Ningqiang Song}
\thanks{Corresponding author.\\ songnq@itp.ac.cn}
\affiliation{Institute of Theoretical Physics, Chinese Academy of Sciences, Beijing, 100190, China}

\author{Yidong Song}
\affiliation{Institute of Theoretical Physics, Chinese Academy of Sciences, Beijing, 100190, China}

\author{Junhua Wang}
\affiliation{Beijing Key Laboratory of Fault-Tolerant Quantum Computing, Beijing Academy of Quantum Information Sciences, Beijing, China}

\author{Yue-Liang Wu}
\affiliation{Institute of Theoretical Physics, Chinese Academy of Sciences, Beijing, 100190, China}
\affiliation{ Taiji Laboratory for Gravitational Wave Universe (Beijing/Hangzhou), University of Chinese Academy of Sciences (UCAS), Beijing 100049, China}
\affiliation{School of Fundamental Physics and Mathematical Sciences,
Hangzhou Institute for Advanced Study, UCAS, Hangzhou 310024, China}
\affiliation{International Centre for Theoretical Physics Asia-Pacific, Beijing/Hangzhou, China}

\author{Tiantian Zhang}
\affiliation{Institute of Theoretical Physics, Chinese Academy of Sciences, Beijing, 100190, China}

\author{Yu-Feng Zhou}
\thanks{Corresponding author.\\ yfzhou@itp.ac.cn}
\affiliation{Institute of Theoretical Physics, Chinese Academy of Sciences, Beijing, 100190, China}
\affiliation{School of Physical Sciences, University of Chinese Academy of Sciences (UCAS), Beijing 100049, China}
\affiliation{School of Fundamental Physics and Mathematical Sciences,
Hangzhou Institute for Advanced Study, UCAS, Hangzhou 310024, China}
\affiliation{International Centre for Theoretical Physics Asia-Pacific, Beijing/Hangzhou, China}

\date{\today}

\begin{abstract}
We present the design and the sensitivity reach of the Qubit-based Light Dark Matter detection experiment. We propose the novel two-chip design to reduce signal dissipation, with quantum parity measurement to enhance single-phonon detection sensitivity. We demonstrate the performance of the detector with full phonon and quasiparticle simulations. The experiment is projected to detect $\gtrsim 30$~meV energy deposition with nearly $100\%$ efficiency and high energy resolution. The sensitivity to $m_\chi\gtrsim 0.01$~MeV  dark matter scattering cross section is expected to be advanced by orders of magnitude for both light and heavy mediators, and similar improvements will be achieved for axion and dark photon absorption in the $0.04$--$0.2$~eV mass range. 
\end{abstract}

\maketitle

\textbf{\textit{Introduction}} ---
The properties of dark matter (DM) have long been one of the outstanding puzzles in fundamental physics. Over the years, multi-tonne scale direct detection experiments have placed stringent constraints on DM above the GeV mass scale~\cite{XENON:2023cxc,LZ:2024zvo,PandaX:2024qfu}, which motivates the search for DM in the sub-GeV mass range with TES- and semiconductor-based experiments~\cite{SuperCDMS:2024yiv,CRESST:2019jnq,SENSEI:2023zdf,DAMIC-M:2023gxo,CDEX:2022kcd}. Theoretical and experimental efforts are undertaken to explore even lower DM mass regimes toward sub-MeV~\cite{Essig:2011nj,Graham:2012su,Essig:2015cda,Hochberg:2015pha,Hochberg:2015fth,Derenzo:2016fse,Hochberg:2017wce,Cavoto:2017otc,Kurinsky:2019pgb,Blanco:2019lrf,Griffin:2020lgd,Essig:2022dfa,Du:2022dxf,Hochberg:2019cyy,Hochberg:2021ymx,Hochberg:2019cyy,Hochberg:2021yud,Das:2022srn,Das:2024jdz,Baudis:2025zyn,Schwemmbauer:2025evp,Baiocco:2025omh,Dutta:2025ddv,Chen:2025cvl,Wu:2025abi}.

For halo DM with velocity $v\sim 10^{-3}$, the energy deposition $\omega$ in scattering scales with the mass of DM $m_\chi$ as $m_\chi v^2$. While MeV--GeV mass DM is heavy enough to ionize the target, sub-MeV DM typically excites only phonons~\cite{Trickle:2020oki}, depositing energy in the sub-eV range below the threshold of conventional direct detection experiments~\cite{XENON:2023cxc,LZ:2024zvo,PandaX:2024qfu,SuperCDMS:2024yiv,CRESST:2019jnq,SENSEI:2023zdf,DAMIC-M:2023gxo,CDEX:2022kcd}. On the other hand, axion~\cite{Peccei:1977ur,Peccei:1977hh,Wilczek:1977pj,Weinberg:1975ui,AxionLimits} and dark photon~\cite{Holdom:1985ag,Dienes:1996zr,Abel:2003ue} can be absorbed by the target and excite a phonon. In the meV mass range, such DM candidates are especially less explored by haloscopes as the design and operation of resonant cavities at the corresponding frequency range remain challenging.

Recent advances at the interface of particle physics and quantum information science provide a promising path forward. Cosmic-ray--induced correlated errors in qubit arrays highlight the possibility of using qubits as the particle detector~\cite{vepsalainen2020impact,wilen2021correlated,martinis2021saving,mcewen2022resolving,Li:2024dpf,harrington2025synchronous}, and qubit-based devices have demonstrated their superior sensitivity to meV photons~\cite{QCD_thoery,QCD_experiment}.

In this \textit{Letter}, we introduce the Qubit-base Light Dark Matter detection experiment carried out at the China Jinping Underground Laboratory (CJPL). In contrast to conceptual studies optimized for on small-scale devices (see, e.g., \cite{Ramanathan:2024hsf,Fink:2023tvb}), we propose a novel two-chip architecture that isolates the DM target from the control lines, employs qubit arrays and quasiparticle traps as single-phonon calorimeters, and reads out quasiparticle tunneling signals using the recently developed Quantum Parity Detector (QPD) technique~\cite{Ramanathan:2024hsf,Li:2024dpf}. This design enables, for the first time, sensitive searches for light DM with both large exposure and high efficiency. Using dedicated simulations, we further develop the full analysis chain from DM scattering to signal processing and show that the proposed detector achieves high efficiency for $\mathcal{O}(10\,\mathrm{meV})$ energy depositions, enabling sensitivity to DM scattering well beyond current limits, while probing both freeze-in and freeze-out DM production scenarios. The sensitivity for axion and dark photon could also be advanced by up to 2--5 orders of magnitude with kg$\cdot$year exposure. 

\begin{figure*}[!htb]
\centering
\includegraphics[width=17cm]{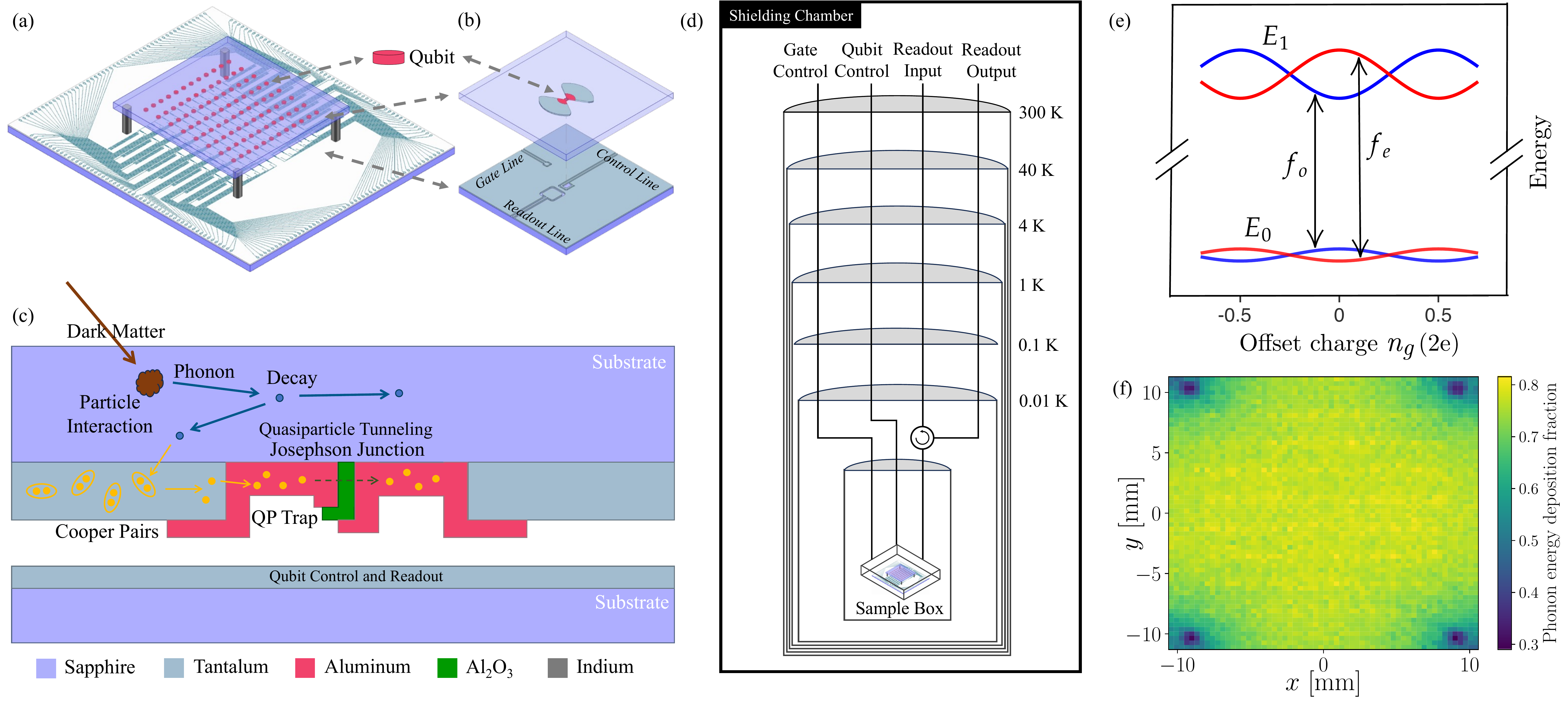}\caption{The schematic drawing of the qubit array and its details. (a) 96 qubits are distributed evenly on the upper slice of the chip, and the lower slice loads the readout, gate and control lines. The 3D structure of a single qubit is enlarged in (b). (c) The qubit response to DM interaction from the lateral view. The materials are labeled with different colors. (d) Lateral view of the full detector. (e) The energy split of the two lowest qubit levels due to even and odd charge parities. (f) The energy fraction of DM-induced phonons absorbed by the qubits as a function of the $x-y$ location of the interaction in the substrate from simulations.}
\label{fig:QPD_figure}
\end{figure*}

\textbf{\textit{Detector design}} ---
The detector employs a flip-chip architecture with a top {\it Qubit chip} and a bottom {\it Carrier chip}, as shown in Figs.~\ref{fig:QPD_figure}(a)--\ref{fig:QPD_figure}(c). The qubit chip with a sapphire substrate hosts 96 qubits, each consisting of a superconducting Al (Aluminum)/Al$_2$O$_3$/Al Josephson junction, and large-area Ta (Tantalum) films galvanically connected to the Al layer serving as capacitor pads. The carrier chip integrates readout resonators, as well as gate and control lines for individual qubits. Despite the dense packing of these lines, we leverage careful structural design to enable approximately 30 dB of isolation between adjacent lines to avoid crosstalk between QPDs~\cite{Kosen:2024cqb}.  DM interacts with the substrate in the qubit chip, producing a phonon that propagates from the substrate to the Ta films, breaking Cooper pairs and creating quasiparticles (QPs). QPs will then be concentrated in the lower-gap Al in the junction. Tunneling of the QPs will change the charge-parity state of the qubit and leads to observable signals. 

To be specific, the parity-dependent Hamiltonian of the qubit
\begin{equation}
    H = 4E_C\left(\hat{n} - n_g + \frac{P-1}{4}\right)^2 - E_J \cos\hat{\phi} ,
\end{equation}
where $E_C$ and $E_J$ are the charge and Josephson energies, $\hat{n}$ and $n_g$ are the number operator for Cooper pairs and the offset charge, $\hat{\phi}$ is the superconducting phase difference across the junction and $P = +1$ ($-1$) labels the even (odd) charge-parity state.  The qubit then possesses two lowest energy levels shown in Fig.~\ref{fig:QPD_figure}(e). Tunneling of a QP will induce a frequency shift $|f_{\mathrm{e}}-f_{\mathrm{o}}|$, which we exploit to map the charge-parity states onto the eigenstates of qubit through a Ramsey-based sequence~\cite{Rist2013,Serniak2018hot}. The qubit array therefore functions as a set of QPDs, capable of resolving individual QP-tunneling events and sensitive to low-energy single phonons. Experimentally, this method has already achieved a time resolution of a few microseconds~\cite{Li:2024dpf,Rist2013}. Moreover, a fidelity of $99.9\%$ for both single-qubit gates and the readout has been demonstrated~\cite{Wang:2024rjw,marxer2025above}. Therefore, we adopt the charge-parity detection fidelity $\mathcal{F}=95\%$ in this work. We also equip each qubit with an additional gate line to compensate the offset charge drift~\cite{Christensen2019Anomalous,wilen2021correlated}, which allows the detector to operate continuously for weeks to months.

Finally, the detector is packaged and operated at the base temperature of a dilution refrigerator around $10~\mathrm{mK}$ (Fig.~\ref{fig:QPD_figure}(d)), and the entire system is enclosed within gamma and neutron shieldings at CJPL. The main low-energy background comes from the tunneling caused by residual QPs. Near-gap infrared photons from the higher temperature stages of the refrigerator may break Cooper pairs and increase the residual QP density $n_{\rm res}$, mainly through the coaxial wiring~\cite{Liu:2022aez}. We apply light filtering in the wires, and use compact sample box coated with infrared-absorbing material to enhance shielding against infrared and stray photons, which is expected to reduce the tunneling rate down to about 0.1 Hz~\cite{Connolly:2023gww,pan2022engineering}, corresponding to the residual QP density $n_{\rm res}\sim 10^{-4}~\mu{\rm m}^{-3}$. Conservatively, we adopt $n_{\rm res}=10^{-3}~\mu{\rm m}^{-3}$ as a benchmark.

As will be demonstrated below, we observe the following distinctive advantages of the detector: (1) The suspended design 
and the qubit array layout
force DM-induced phonon to dissipate to the environment (carrier chip) only through the four In (indium) pillars in the corners. Simulations suggest that more than 80\% of the phonon energy is absorbed in the qubits.  (2) The Ta-Al design will concentrate QPs in the Al trap, which significantly increases the QP tunneling rate. (3) The detector is sensitive to single-QP tunneling and low-energy single phonons. The simulated detection efficiency is close to $100\%$ in the sub-eV range, in contrast to $\mathcal{O}({\rm eV})$ threshold in TES- and skipper-CCD-based DM experiments~\cite{SuperCDMS:2024yiv,SENSEI:2023zdf}. We also reach a high energy resolution of about 50\% at 100~meV, allowing to reject higher-energy backgrounds. (4) The detector mass is scalable with little loss of the detection efficiency. We propose different geometries for the qubit chip---a ``thin chip'' of 0.82~g ($21.2\times 22.6\times 0.43$~mm) typical for quantum computation~\cite{Li:2024dpf} and a ``thick chip'' of 38~g ($21.2\times 22.6\times 20$~mm). The final phase of the experiment will be kilogram-mass with arrays of thick chips.

\textbf{\textit{DM interaction rate}} ---
Light DM $\chi$ may scatter with the electrons in the substrate and excite a sub-eV phonon, at a rate related to the response of the material. The differential scattering rate $\mathrm{d}R/\mathrm{d}\omega$ is computed with the \texttt{PhonoDark}~\cite{Trickle:2020oki} code. As inputs, we calculate the phonon spectra, equilibrium positions of the ions, the potential energy, and effective ionic charges using first-principle density functional theory with the \texttt{VASP} code~\cite{kresse1996efficient,kresse1996efficiency}.

The detector can also be used to detect DM absorption. The absorption rate of dark photon in the sapphire substrate is proportional to the imaginary part of the reciprocal of the dielectric function~\cite{Knapen:2021bwg,Berlin:2023ppd,Hochberg:2016sqx,Griffin:2018bjn,Mitridate:2021ctr,Mitridate:2023izi}, which we extract from the \texttt{darkELF} code~\cite{Knapen:2021bwg}. The energy deposition is nearly monochromatic in the form of a phonon at the mass of dark photon. Axion could be detected similarly by applying a strong magnetic field.

For both DM scattering and absorption, the expected number of signal events in the detector of mass $M$ during operation time $\mathcal{T}$ is
\begin{equation}
    N_{\rm sig}=\int \mathrm{d}\omega M\mathcal{T}\dfrac{\mathrm{d}R}{\mathrm{d}\omega}\int \mathrm{d}\omega_{\rm rec}\varepsilon(\omega)f_{\rm r}(\omega_{\rm rec},\omega) \,,
    \label{eq:Nsig}
\end{equation}
where $\omega_{\rm rec}$ is the reconstructed phonon energy, and $\varepsilon$ and $f_{\rm r}$ are the efficiency and the resolution functions.

\textbf{\textit{Phonon transport and quasiparticle tunneling rate}} ---
 We simulate propagation and full dynamics of the phonons from DM interaction with the \texttt{G4CMP} code~\cite{Kelsey:2023eax} using the designed geometry of the qubit detector. The phonons are uniformly injected in the substrate at energy $\omega$, with an initial polarization state sampled from the partition fractions of the possible polarization states. The phonons will scatter with lattice defects such as isotopes to change their direction and polarization states, at a rate $B\nu^4$, and the longitudinal phonons may decay at a rate $A\nu^5$. Upon reaching the sapphire-Ta interface, phonons have a probability of 94\% to transmit into the Ta absorber computed from the Acoustic Mismatch Model~\cite{Swartz1989rmp}, and hardly escape back to the substrate once absorbed~\cite{Kelsey:2023eax}. Unabsorbed phonons are reflected assuming specular reflection as investigated in~\cite{Martinez:2018ezx}. Phonons may be lost by exciting surface modes when interacting with imperfectly smooth substrate boundaries. Following  CDMS~\cite{CDMSsimulation}, we assume a single-pass surface loss probability of 0.1\%. 

 Fig.~\ref{fig:QPD_figure}(f) shows the fraction of phonon energy absorbed by the qubits (rather than the In pillars) as a function of the $x-y$ location of DM interaction. The average fraction is more than 80\% in the majority of the substrate with the multi-qubit design, as opposed to  $\lesssim 10\%$ for similar quantum devices~\cite{Martinez:2018ezx,Linehan:2025suv}. Near the pillars the fraction drops to below 30\%.

Phonons entering the Ta absorber with energy $E_{\rm ph}>2\Delta_{\rm abs}$, twice the superconducting gap of the absorber, will be able to break Cooper pairs in the Kaplan cascade process~\cite{Kaplan1976prb} and produce QPs approaching $\Delta_{\rm abs}$. The average fraction of the energy released to QPs is quantified by $\eta_{\rm pb}$, which includes the energy loss to sub-gap phonons.
We adopt $\eta_{\rm pb}=0.6$ as suggested by the literature~\cite{guruswamy2014quasiparticle,kozorezov2000quasiparticle}.

Since the superconducting gap of the Al trap $\Delta_{\rm trap}$ is substantially lower than the Ta absorber,  QPs will then transport from the absorber to the trap. We introduce the trapping efficiency $\eta_{\rm trap}$ to quantify the fraction of QPs in the absorber that are transmitted to and finally bound in the trap. We study the trapping process by solving the diffusion equation of QPs in between the absorber and the trap, showing $\eta_{\rm trap}\sim 0.7$ in our setup. Hence the number density of QPs in the trap due to the absorption of phonon energy $E_{\rm ph}$ in the absorber is $n_{\rm qp}=\eta_{\rm pb}\eta_{\rm trap} E_{\rm ph}/V_{\rm trap}\Delta_{\rm trap}$, where $V_{\rm trap}$ is the volume of the trap.

After creation, the QPs may deplete exponentially in the recombination process at a time scale $\tau_{\rm qp}$ of a few ms~\cite{barends2009enhancement,wang2014measurement}. Conservatively we take $\tau_{\rm qp}=1$~ms in this work. The actual number of QPs in the trap may also deviate from expectation due to the correlation in the QP creation process captured by the Fano factor $F\sim 0.2$~\cite{verhoeve2002superconducting}, which then follows a normal distribution with a width $\sigma=\sqrt{Fn_{\rm qp}V_{\rm trap}}$~\cite{fano1947ionization}.
The tunneling rate is computed to be~\cite{palmer2007steady,shaw2008kinetics,shaw2009quantum,lutchyn2007kinetics}
\begin{equation}
    \Gamma_{\rm tun}(t)\simeq \dfrac{16E_J k_B T}{h\mathcal{N}_{\rm qp}\Delta_{\rm trap}}n_{\rm qp}e^{-t/\tau_{\rm qp}}\,,
    \label{eq:tunnelingrate}
\end{equation}
in the low temperature limit $T\ll \Delta_{\rm trap}$, where $\mathcal{N}_{\rm qp}$ represents the number of quasiparticle states available in the trap. 
The background tunneling rate due to non-zero residual QP density $n_{\rm res}$ in the trap can be calculated similarly using Eq.~\eqref{eq:tunnelingrate}, by replacing $n_{\rm qp}e^{-t/\tau_{\rm qp}}$ with $n_{\rm res}$.

The setup, relevant parameters and the effects of different residual QP density are detailed in the Supplemental Material~\cite{supp}.

\textbf{\textit{Detector resolution and efficiency}} ---
The number of tunnelings at a qubit in one measurement is the sum of phonon-induced tunnelings $N_{s}$ and residual QP-induced tunnelings $N_{0}$, which we draw from the Poisson distribution, considering the detection fidelity at the same time ((we design a time-domain filtering process to suppress readout-induced dark counts to two orders of magnitude below the residual quasiparticle tunneling rate, see the Supplemental Material~\cite{supp} for details.).

\begin{figure}[!htb]
    \centering
    \includegraphics[width=\columnwidth]{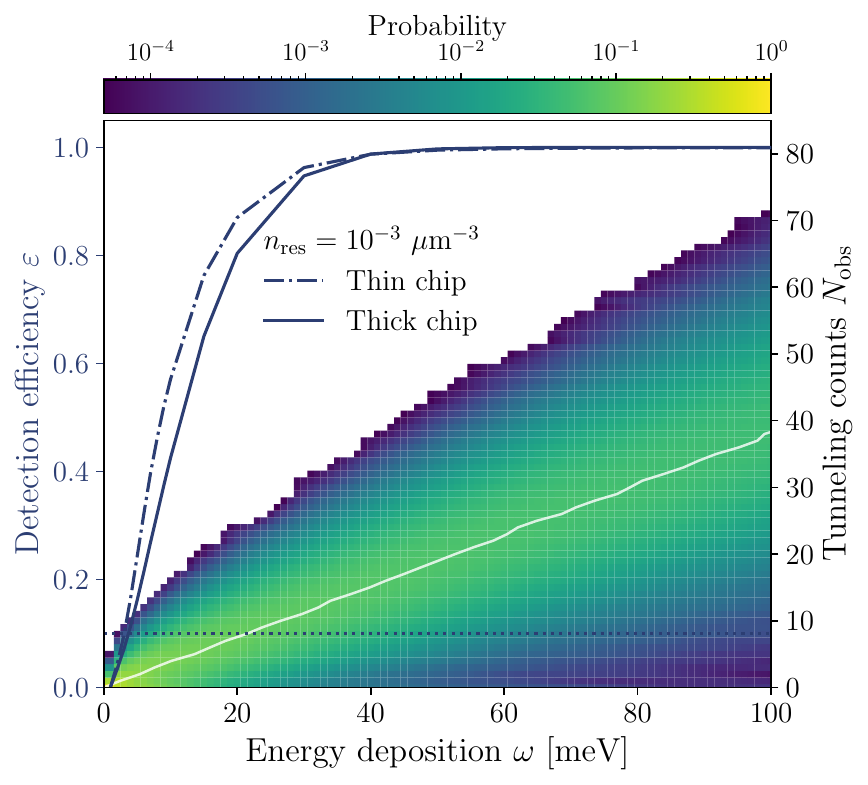}
    \caption{The blue lines show the detection efficiency as a function of the energy deposition from DM interaction for a thin-chip (dash-dotted) and a thick-chip (solid) detector. The dotted line marks the 10\% efficiency. The colors depict the probability distribution of tunneling counts in all qubits in one measurement time window $N_{\rm obs}$ for a thin chip as a function of the energy deposition, and the white line marks the reconstructed energy corresponding to the maximum likelihood as a function of $N_{\rm obs}$.}
    \label{fig:efficiency}
\end{figure}

We identify the DM-induced phonon energy $\omega$ from the observation of the total number of tunnelings $N_{\rm obs}$ of all qubits in a 1.5 ms time window.
 The reconstructed energy can be obtained using maximum likelihood method, i.e., $\omega_{\rm rec} = {\max}_{\omega}\,\mathcal{L}(N_{\rm obs}|\omega)$, where the likelihood $\mathcal{L}$ is constructed from the simulations as shown in Fig.~\ref{fig:efficiency}. The energy resolution, defined as the ratio of full width at half maximum (FWHM) and $\omega_{\rm rec}$, decreases for larger phonon energy and reaches about 50\% at 100~meV. The energy resolution function $f_r$ is constructed as Gaussian with $\sigma=\textrm{FWHM}/2.355$. Therefore, high energy background events falling outside the detection window can be rejected.

We now find the detection efficiency given the background tunnelings from residual QPs. To identify a signal, we require that the total number of tunnelings for $N_{\rm qb}=96$ qubits in a measurement $j$ to deviate from the background tunneling $N_b=\mathcal{F}N_{0}N_{\rm qb}$ (background-only hypothesis) at least at $3\sigma$ C.L. assuming $\chi^2$ distribution, i.e., $\Sigma_{N_{\rm obs}=N_j}^{\infty} \mathrm{Poisson}(N_{\rm obs},\mu=N_b)<0.0027$. We then find the efficiency as the fraction of simulations that satisfy this condition. 

The detection efficiency $\varepsilon$ is displayed in Fig.~\ref{fig:efficiency}, which rapidly approaches 100\% as the phonon energy increases. 10\% efficiency corresponds to $\omega\sim 3$~meV both for thin- and thick-chip designs, orders of magnitude lower than the threshold of typical DM direct detection experiments. The thick-chip design degrades the efficiency only slightly compared to a thin chip.  Although phonons travel longer in the thick-chip substrate and tend to decay more to lower energy phonons below the pair-breaking threshold, few-meV phonons have mean free path exceeding the size of the thick chip, preventing efficient decay.

\begin{figure*}[!htb]
    \centering
    \includegraphics[width=\columnwidth]{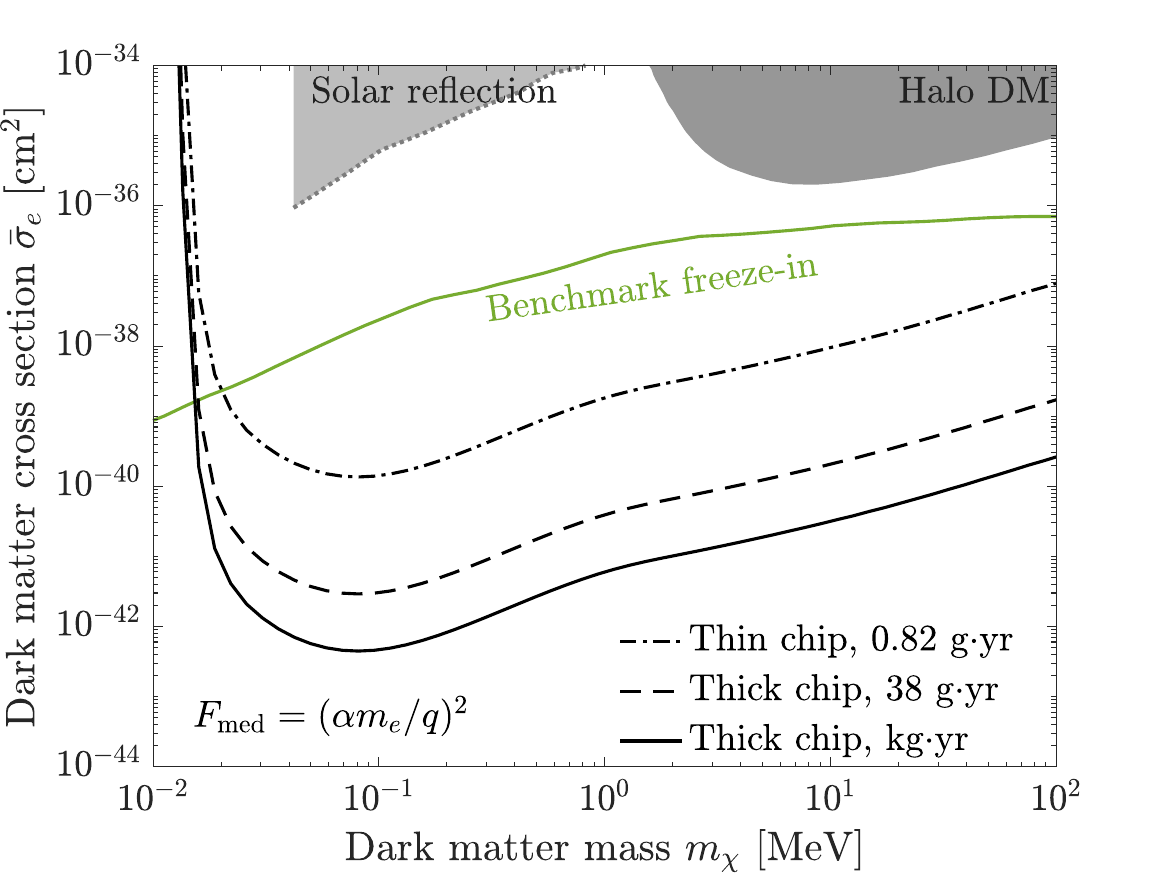}
    \includegraphics[width=\columnwidth]{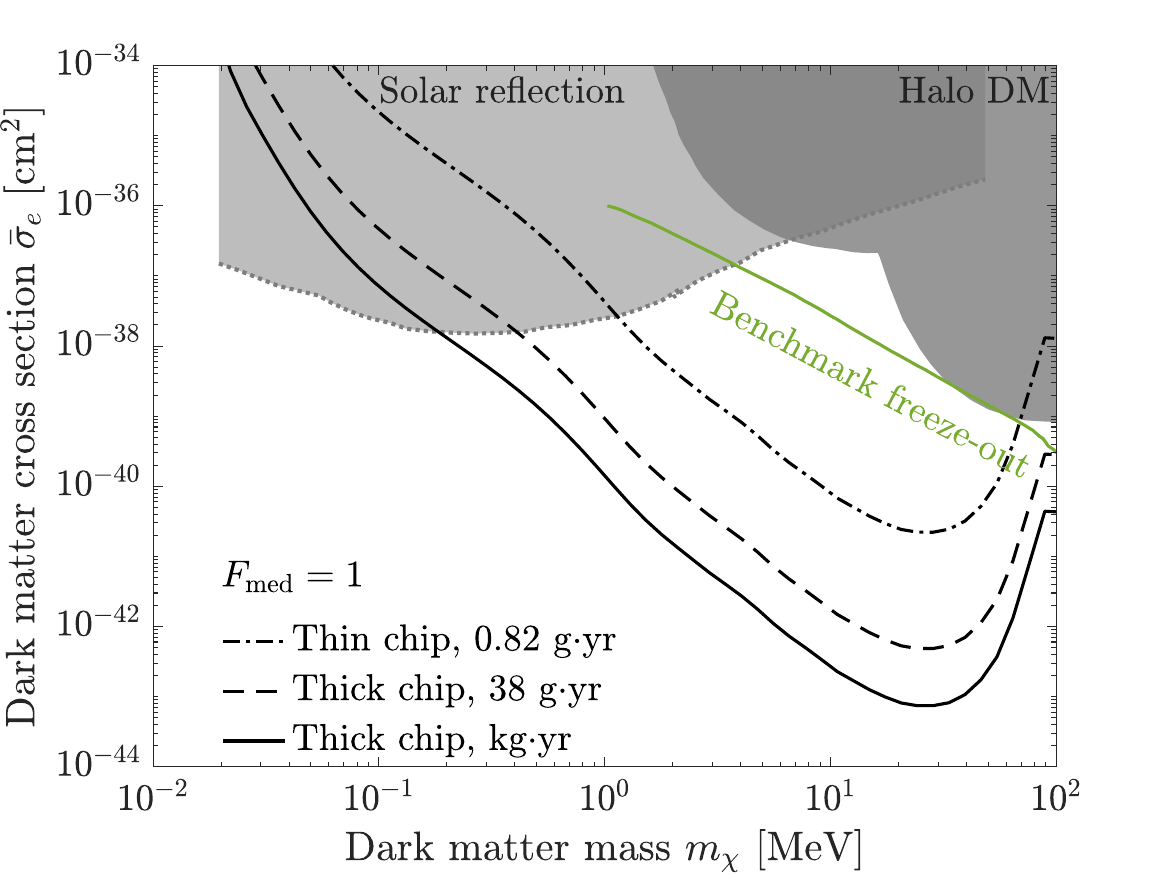}
    \includegraphics[width=\columnwidth]{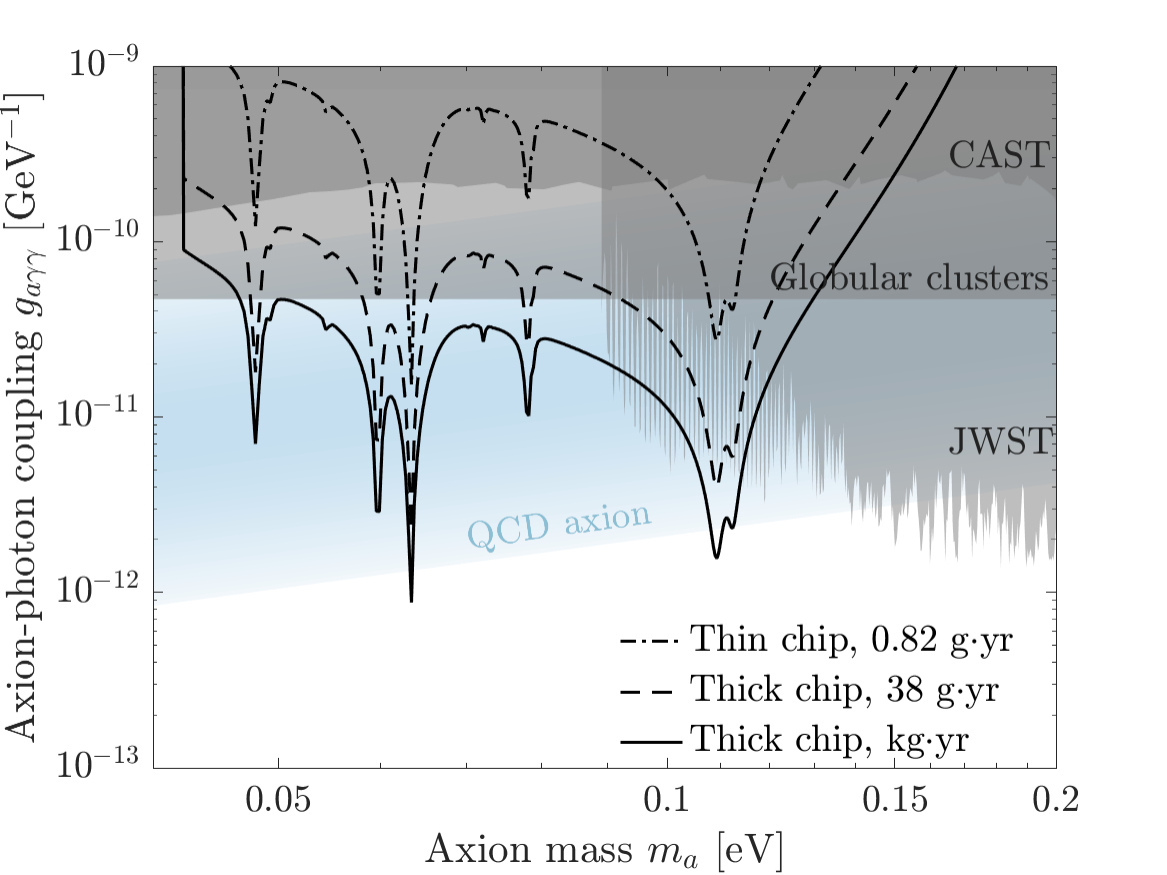}
    \includegraphics[width=\columnwidth]{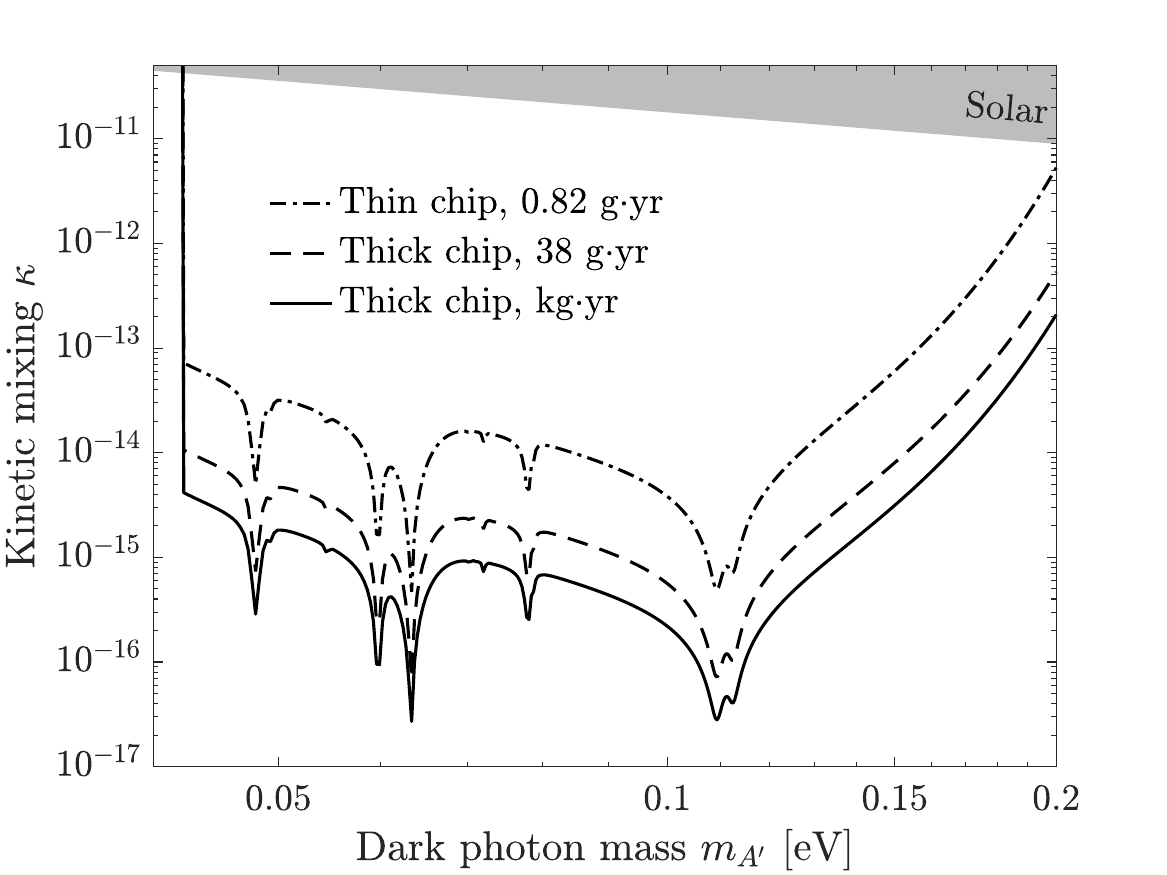}       
    \caption{\textit{Upper:} 90\% C.L. sensitivity on DM-electron scattering cross section. Black lines correspond to different exposures using the detector design in this work. We also show existing constraints from the direct detection of halo DM~\cite{SENSEI:2023zdf,DAMIC-M:2023gxo,DarkSide:2022knj,XENON:2019gfn,EDELWEISS:2020fxc,PandaX-II:2021nsg,SuperCDMS:2024yiv}, and the DM accelerated and reflected by electrons in the Sun~\cite{An:2021qdl,Emken:2024nox}. The green lines show the cross section to produce the correct DM relic density in the benchmark freeze-in process~\cite{Essig:2011nj,Chu:2011be,Essig:2015cda,Dvorkin:2019zdi} for a light mediator (left) and freeze-out process~\cite{Essig:2015cda,Boehm:2003hm,Lin:2011gj,Hochberg:2014dra,DAgnolo:2019zkf,Kahn:2021ttr} for a heavy mediator (right). \textit{Lower:} 90\% C.L. sensitivity on DM absorption. For axion we also show existing constraints on axion-photon coupling $g_{a\gamma\gamma}$ from CAST~\cite{CAST:2017uph,CAST:2024eil}, the globular clusters~\cite{Ayala:2014pea,Dolan:2022kul} and JWST~\cite{Pinetti:2025owq,Saha:2025any} (left). The blue band shows parameter space for QCD axion~\cite{Peccei:1977ur,Peccei:1977hh,Wilczek:1977pj,Weinberg:1975ui,AxionLimits}. For dark photon we show limits on kinetic mixing $\kappa$ from solar cooling~\cite{Vinyoles:2015aba,An:2013yfc,Li:2023vpv} (right).}
    \label{fig:Al2O3sensitivity}
\end{figure*}

\textbf{\textit{Sensitivity and background}} ---
We derive the projected limits on DM parameters by requiring the number of signal events from DM interaction computed in Eq.~\eqref{eq:Nsig} not to exceed the expected background at 90\% C.L. We only analyze the single-phonon regime where the energy $\omega_{\rm rec}\leq 100$~meV, and multi-phonon and ionization signals will be explored in future work.

We incorporate two types of background---tunnelings caused by residual QPs and the soft scattering of high energy photons, with the former accounted for in the detection efficiency. Cosmic rays and radioactive decay typically lead to energy deposition much higher than keV away from the signal region and are largely shielded by the rock and shielding at CJPL. However, high energy photons produced inside the refrigerator may scatter softly and deposit energy below 100~meV, which is indistinguishable from DM interaction on event-by-event basis. We adopt the background calculation in~\cite{Berghaus:2021wrp}, using the benchmark background photon spectrum measured by the EDELWEISS Collaboration~\cite{EDELWEISS:2018tde}. This amounts to the deployment of 20~cm of lead and 50~cm of polyethylene to shield against gamma and neutrons in our setup. The expected background rate is 5.27~kg$^{-1}\cdot$yr$^{-1}$, which we use as a benchmark in this work.

In addition, the stress released due to the difference in the thermal contraction of the detector materials during the cooling phase may also cause phonon bursts in the qubit detector~\cite{anthony2024stress}. The stress-induced background decreases exponentially in a typical time scale of 6--10 days~\cite{anthony2024stress,Yelton:2025wsy} and could be eliminated by waiting long enough time before measurement. Therefore, we do not include them in the calculation.

\revAdd{We note that low energy excess (LEE) above eV has been observed in several experiments using different technologies (see~\cite{Baxter:2025odk} for a review). The sub-100 meV background targeted by our qubit detector may or may not share the same origin as LEE. The background in this specific energy range will be measured and characterized upon experimental operation, and the ultimate sensitivity to scattering cross-sections will scale proportionally with this background level. We refer to the Supplemental Material~\cite{supp} for more detailed discussions on the anticipated origins and mitigation strategies for these backgrounds.}

We show the projected sensitivity on the DM-electron scattering cross section in the upper panels in Fig.~\ref{fig:Al2O3sensitivity}. For both light and heavy mediators, the projected limits of a single thin-chip detector with 0.82 g$\cdot$yr exposure surpass the existing constraints by orders of magnitude, including the limits from the direct detection of halo DM~\cite{SENSEI:2023zdf,DAMIC-M:2023gxo,DarkSide:2022knj,XENON:2019gfn}, and DM reflected from the sun~\cite{An:2021qdl,Emken:2024nox}, allowing to examine the benchmark freeze-in~\cite{Essig:2011nj,Chu:2011be,Essig:2015cda,Dvorkin:2019zdi} and freeze-out~\cite{Essig:2015cda,Boehm:2003hm,Lin:2011gj,Hochberg:2014dra,DAgnolo:2019zkf,Kahn:2021ttr} scenarios.

We also show the sensitivity on dark photon kinetic mixing, which improves over  the constraints from solar cooling~\cite{Vinyoles:2015aba,An:2013yfc,Li:2023vpv} by 4--5 orders of magnitude  with a single thin qubit chip. A similar process could be leveraged to detect axion DM with the application of strong magnetic field $B_0$, the feasibility of which has recently been demonstrated experimentally in~\cite{Gunzler:2025spt}. 
A single thin chip with one year running can already exclude new parameter space with $B_0=10$~T, and kg$\cdot$yr exposure can advance the current constraints from globular clusters by up to about two orders of magnitude.

\textbf{\textit{Conclusions and prospects}} ---
We demonstrate the performance of a newly designed cryogenic qubit-array detector. Although we focus on the sensitivity to DM-electron scattering and DM absorption, similar sensitivity and mass reach could be achieved for DM coupling to nucleons~\cite{Trickle:2019nya}. 

In this study, we use only the total number of observed tunnelings to reconstruct the phonon energy. In a multi-qubit setup, topology of tunneling qubits and timing correlations could further improve energy and position reconstruction, as well as signal--background separation---since phonon-induced tunnelings decay rapidly in time while residual tunneling events remain nearly constant. Moreover, reducing $\Delta_{\rm trap}$ could further enhance the detection efficiency, while changing the substrate material will modify the detector's response to different DM models~\cite{Griffin:2019mvc,Campbell-Deem:2022fqm}.  A detailed exploration of these improvements is deferred to future work.

\textbf{\textit{Acknowledgements}} ---
We thank Elena Aprile, Suerfu Burkhant, Zhaozheng Lv, Yifan Chen, and Bin Zhu for useful discussions. N.S. is supported by the National Natural Science Foundation of China (NSFC) Project Nos. 12441504, 12347105, 12475110 and 12447101. J.S. is supported by Peking University under startup Grant No. 7101302974 and the NSFC under Grants No. 12025507, No. 12450006. J.W. and Y.Z. are supported by the NSFC under the Project No. 12441504. T.Z. acknowledges the support from National Key R\&D Project Grant Nos. 2023YFA1407400 and 2024YFA1409200, and the NSFC Grant Nos. 12374165 and 12447101.

\bibliography{qubitdm}

\newpage
\clearpage
\onecolumngrid

\setcounter{equation}{0}
\setcounter{figure}{0}
\setcounter{section}{0}
\setcounter{table}{0}
\setcounter{page}{1}
\renewcommand{\theequation}{S.\arabic{equation}}
\renewcommand{\thefigure}{S\arabic{figure}}
\renewcommand{\thetable}{S\arabic{table}}
\renewcommand{\thesection}{\Roman{section}}
\newcommand{\fakeaffil}[2]{$^{#1}$\textit{#2}\\}

\begin{center}
\vspace{0.5in}
\textbf{\large Novel Light Dark Matter Detection with Quantum Parity Detector Using Qubit Arrays}
\vspace{0.05in}
\\
{ \it \large Supplemental Material}\\ 
\medskip
{Xuegang Li,\textsuperscript{1} Yuxiang Liu,\textsuperscript{2,1} Jing Shu,\textsuperscript{2,3,4} Ningqiang Song,\textsuperscript{5} \\Yidong Song,\textsuperscript{5} Junhua Wang,\textsuperscript{1}
Yue-Liang Wu,\textsuperscript{5,6,7,8} Tiantian Zhang,\textsuperscript{5} and Yu-Feng Zhou\textsuperscript{5,9,7,8}} \par\medskip
{\small  \fakeaffil{1}{Beijing Key Laboratory of Fault-Tolerant Quantum Computing,\\ Beijing Academy of Quantum Information Sciences, Beijing, China}
 \fakeaffil{2}{School of Physics and State Key Laboratory of Nuclear Physics and Technology, Peking University, Beijing 100871, China}
 \fakeaffil{3}{Center for High Energy Physics, Peking University, Beijing 100871, China}
 \fakeaffil{4}{Beijing Laser Acceleration Innovation Center, Huairou, Beijing, 101400, China}
 \fakeaffil{5}{Institute of Theoretical Physics, Chinese Academy of Sciences, Beijing, 100190, China}
 \fakeaffil{6}{Taiji Laboratory for Gravitational Wave Universe (Beijing/Hangzhou),\\ University of Chinese Academy of Sciences (UCAS), Beijing 100049, China}
 \fakeaffil{7}{School of Fundamental Physics and Mathematical Sciences,\\
Hangzhou Institute for Advanced Study, UCAS, Hangzhou 310024, China}
\fakeaffil{8}{International Centre for Theoretical Physics Asia-Pacific, Beijing/Hangzhou, China}
\fakeaffil{9}{School of Physical Sciences, University of Chinese Academy of Sciences (UCAS), Beijing 100049, China}}
\end{center}


\vspace{3mm}


In this material, we present details of the dark matter (DM) scattering rate, the full setup of the detector and the readout scheme, values for relevant parameters used in this work, the phonon and quasiparticle (QP) transport processes, the QP tunneling rate, and relevant backgrounds.

\section{Details of the density functional theory calculations}
\label{sec:DFT}

Assuming DM interacts with electrons, the scattering rate per unit of target mass is~\cite{Griffin:2019mvc,Trickle:2019nya,Trickle:2020oki}
\begin{equation}
    R=\dfrac{\rho_\chi}{m_{\rm cell}m_\chi}\dfrac{\pi\bar{\sigma}_e}{2\mu_{\chi e}^2}\int\dfrac{\mathrm{d}^3\bm{q}}{(2\pi)^3}F^2_{\rm med}(q)\sum\limits_\nu \dfrac{1}{\omega_{\nu,\bm{k}}}\times\left\lvert\sum\limits_j\dfrac{e^{-W_j}}{\sqrt{m_j}}e^{i\bm{G}\cdot\bm{x}_j^0}\bm{Y}_j\cdot\bm{\epsilon}^*_{\nu,\bm{k},j}\right\rvert^2g(\bm{q},\omega_{\nu,\bm{k}})\,,
    \label{eq:scatteringrate}
\end{equation}
where $m_{\rm cell}=\rho_T\Omega$ is the mass contained in the primitive cell and $\Omega$ is the volume of the primitive cell. DM must excite the phonon eigenmodes $\omega_{\nu,\bm{k}}$, with $\nu$ the phonon branch and $\bm{k}=\bm{q}+\bm{G}$ the phonon momentum, and $\bm{G}$ is the reciprocal latter vector. $\bm{\epsilon}_{\nu,\bm{k},j}$ are the phonon polarization vectors with $j$ labeling the atoms in the primitive cell. $W_j$ is the material-specific Debye-Waller factor and $Y_j$ describes how DM interacts with the material. $g$ evaluates the DM velocity integral in the phase space capable of exciting the phonon mode.

The scattering rate also depends on the properties of DM. In Eq.~\eqref{eq:scatteringrate}, $\bar{\sigma}_e$ is the DM-electron scattering cross section at the reference momentum transfer $q_0=\alpha m_e$ and $\mu_{\chi e}$ is the reduced mass between DM and electron. The scattering rate also depends on mediator mass where $F_{\rm med}=q_0^2/q^2$ for a light mediator and $F_{\rm med}=1$ for a heavy mediator. The factor $g$ encapsulates the DM velocity integral.

\begin{figure}[!ht]
    \centering
    \includegraphics[width=0.5\columnwidth]{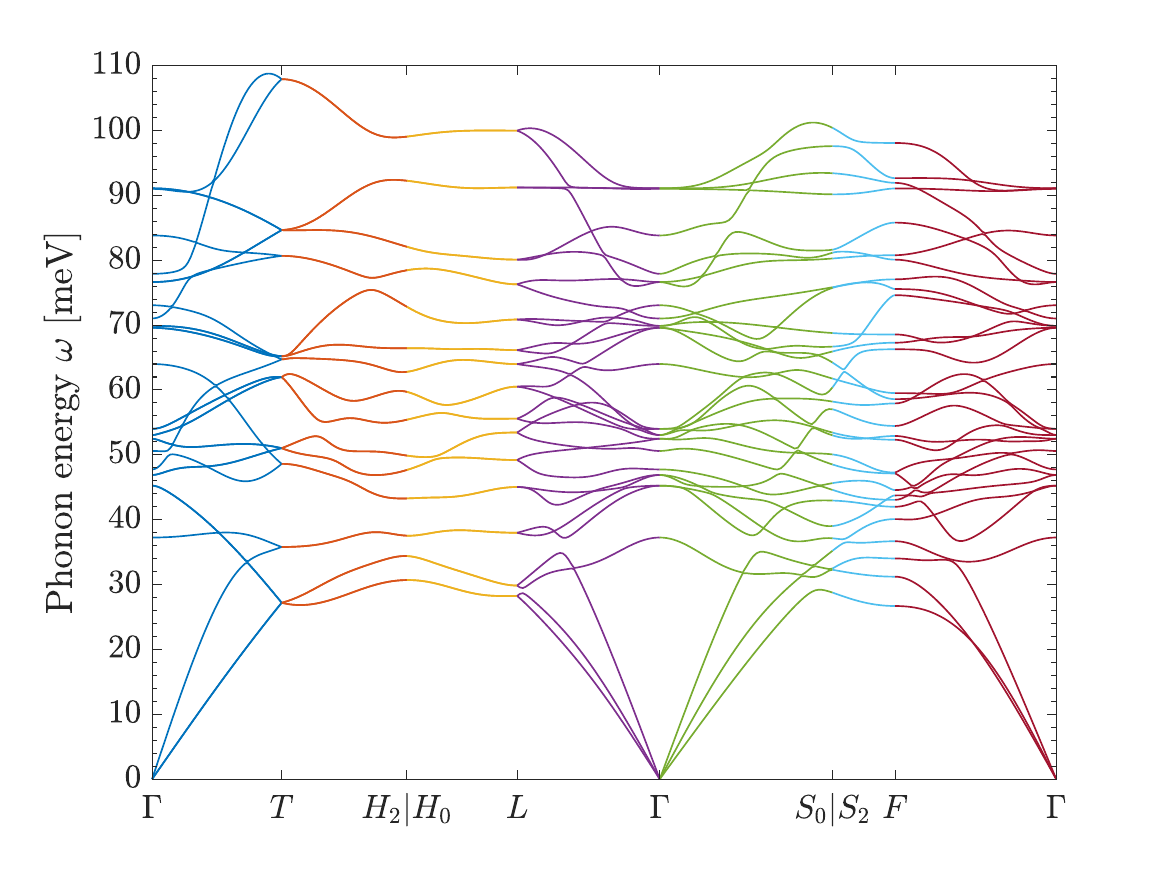}
    \caption{The phonon spectra of sapphire computed using DFT with the \texttt{VASP} code. The letters in the $x$-axis label the high symmetry points in the Brillouin zone. 
   Different colors mark the paths between different points.}
    \label{fig:phononbands}
\end{figure}

We use sapphire (Al$_2$O$_3$) as the target for DM scattering, and calculate the phonon spectra using the Vienna Ab Initio Simulation Package (VASP)~\cite{kresse1996efficient,kresse1996efficiency} within the framework of density functional theory (DFT). The calculations employ a conventional cell with lattice parameters \(a = b = 9.55~\text{\AA}\) and \(c = 13.01~\text{\AA}\). A \(2 \times 2 \times 2\) supercell is constructed relative to the conventional cell, and the phonon dispersions are obtained using the finite-displacement method.
A plane-wave basis set with a kinetic energy cutoff of 500 eV is used, and Brillouin zone integration is performed on a \(3 \times 3 \times 1\) Monkhorst--Pack $k$-point mesh. The exchange--correlation functional is treated within the generalized gradient approximation, using the Perdew--Burke--Ernzerhof functional~\cite{perdew2008restoring}. 
The sapphire phonon spectra from  our DFT calculations are displayed in Fig.~\ref{fig:phononbands}. 
The labels on the horizontal axis correspond to the positions of high-symmetry points in the Brillouin Zone.

With the DFT inputs, we compute the differential scattering rate $\mathrm{d}R/\mathrm{d}\omega$ with the \texttt{PhonoDark}~\cite{Trickle:2020oki} code, which is also consistent with the results in~\cite{Knapen:2021bwg}.

The detector can also be used to detect DM absorption. For example, the absorption rate of dark photon in the sapphire substrate is related to the dielectric function $\epsilon$ via the relation~\cite{Knapen:2021bwg,Berlin:2023ppd,Hochberg:2016sqx,Griffin:2018bjn,Mitridate:2021ctr,Mitridate:2023izi}
\begin{equation}
    R=\dfrac{1}{\rho_T}\dfrac{\rho_{A'}}{m_{A'}}\kappa^2 m_{A'}\mathrm{Im} \left[\dfrac{-1}{\epsilon(m_{A'})}\right],
    \label{eq:ratedarkphoton}
\end{equation}
where $\kappa$ is the kinetic mixing between the Standard Model photon and the dark photon, and $\rho_{A'}$ and $m_{A'}$ are the energy density and mass of dark photon. The energy deposition $\omega\simeq m_{A'}$ is nearly monochromatic in the form of a phonon, and the momentum transfer $q/\omega\ll 1$. We extract the dielectric function of sapphire from the \texttt{darkELF} code~\cite{Knapen:2021bwg}. Axion could be converted to photon (or electric field) in the magnetic field $B_0$, which is then absorbed and excites a phonon. The corresponding rate is computed from Eq.~\eqref{eq:ratedarkphoton} by replacing $\kappa m_{A'}$ with $g_{a\gamma\gamma}B_0$~\cite{Berlin:2023ppd}.

\section{Detector Operating Principle}
\label{sec:detectorprinciple}

The qubit and carrier chips in our detector are mechanically connected only via indium pillars located at the four corners. The substrate in the qubit chip, made of high-quality polished sapphire (with silicon or other materials as alternatives), thus serves as the target for DM interaction and phonon production. We neglect DM direct interaction with the superconducting films directly, as the mass of the films is much smaller than the substrate. To detect the resulting phonon signal, we fabricate an array of transmon superconducting qubits on the backside of the qubit chip. On the topside of the carrier chip, we pattern Ta (Tantalum) films to form control, gate, and readout lines for manipulating and measuring the qubit states. This suspended absorber geometry allows phonons to be captured efficiently by the qubits and significantly slows their dissipation into the environment. At the same time, the carrier chip remains in good thermal contact with its surroundings, keeping the temperature of the whole device as low as possible and suppressing background phonons. Next, we will first derive the detector's Hamiltonian, and then discuss the detector's operating principle. More details on Josephson junction-based quantum devices can be found in reviews~\cite{Makhlin2001Quantum,Blais2021Circuit,Krantz2019quantum}.

\begin{figure}[!h]
    \centering
    \includegraphics[width=1\columnwidth]{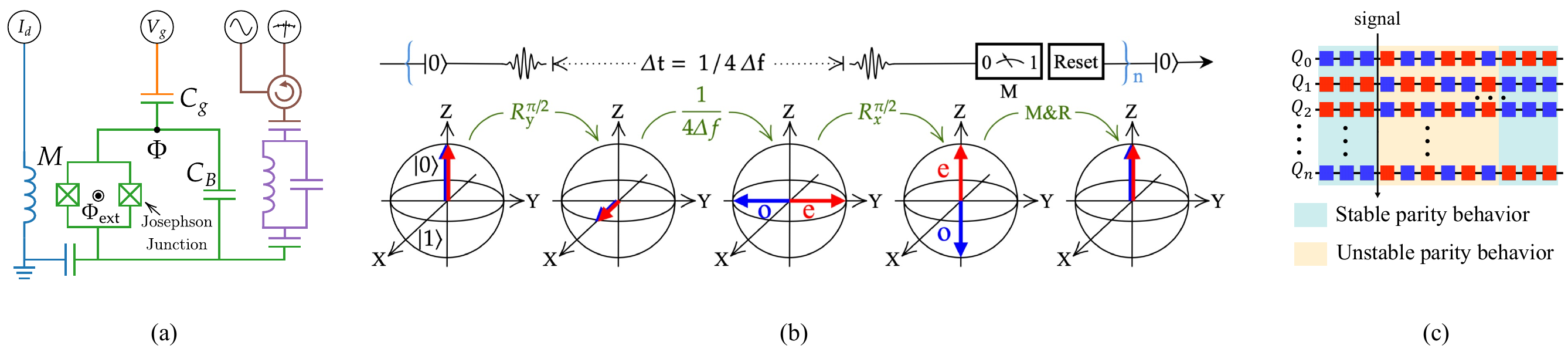}
    \caption{{(a)} Schematic illustration of the qubit circuit.  The qubit (green) comprises a Superconducting Quantum Interference Device (SQUID) and a large shunted capacitance $C_B$, where $\Phi$ is the node flux. The SQUID can be tuned by the external flux $\Phi_\text{ext}$. This qubit can be controlled by a gate voltage $V_g$ (yellow) through capacitance $C_g$ and current drive $I_d$ (blue) through mutual inductance, respectively. To realize a qubit readout, we capacitively couple a LC resonator to the qubit and followed by a readout transmission chain (brown). (b) The charge-parity detection process. The qubit state and charge-parity state at each step of the circuit are depicted on a Bloch sphere. (c) The behavior of a qubit array when capturing a signal.}
    \label{fig:setup}
\end{figure}

The transmon qubit is typically designed for exponential suppression of charge noise sensitivity. In this work, we use an offset-charge-sensitive transmon, intentionally preserving a significant dependence of its transition frequency on the offset charge. The qubit consists of two superconducting capacitor pads connected via a Superconducting Quantum Interference Device (SQUID), comprising two Josephson junctions in parallel. To fully control the qubit, we capacitively couple the qubit to the gate voltage control line via $C_g$ and inductively couple the qubit to the current drive line via $M$, as shown in the circuit of Fig.~\ref{fig:setup}(a). The Lagrangian is:
\begin{equation}
\mathcal{L} = T - U = \frac{1}{2}C_{B}\dot{\Phi}^2 + \frac{1}{2}C_g(\dot{\Phi} - V_g(t))^2 - U_J(\Phi,\Phi_{\text{ext}}) +\frac{M}{L_J}I_d\Phi,
\end{equation}
where $U_J(\Phi,\Phi_{\text{ext}})$ is the Josephson potential energy of the SQUID and $L_J$ is the Josephson inductance of the SQUID.
The conjugate momentum of $\Phi$ is:
\begin{equation}
Q = \frac{\partial \mathcal{L}}{\partial \dot{\Phi}} = C_\Sigma\dot{\Phi} - C_g V_g(t),
\end{equation}
where $C_\Sigma = C_B + C_g$ is the total capacitance. The Hamiltonian is then obtained via Legendre transformation:
\begin{equation}
H = Q\dot{\Phi} - \mathcal{L} = \frac{Q^2}{2C_\Sigma} + \frac{C_g}{C_\Sigma} V_g(t) Q - \frac{M}{L_J}I_d(t)\Phi +U_J(\Phi,\Phi_{\text{ext}}) + \text{const}.
\end{equation}
The constant term is independent of the dynamical variables $Q$ and $\Phi$ and can be omitted. We introduce the dimensionless variables
\begin{equation*}
    n = \frac{Q}{2e},\quad
n_g(t) = -\frac{C_g V_g(t)}{2e}, \quad
\phi = \frac{2\pi\Phi}{\Phi_0},\quad
E_C = \frac{e^2}{2C_\Sigma},
\end{equation*}
where $\Phi_0 = h/2e$ is the flux quantum. The inductive coupling term becomes $-\xi I_d(t)\phi$ with strength $\xi = \frac{M}{L_J}\frac{\Phi_0}{2\pi}$. Upon quantization, $\phi$ and $n$ become operators satisfying $[\hat{\phi}, \hat{n}] = i$, leading to the Hamiltonian
\begin{equation}\label{eq:Hamiltonian}
H = 4E_C \bigl(\hat{n} - n_g(t)\bigr)^2 + U_J(\hat{\phi},\Phi_{\text{ext}}) - \xi I_d(t)\hat{\phi}.
\end{equation}
The Josephson potential energy for a symmetric SQUID~\cite{Koch2007Charge} is:
\begin{equation}\label{eq:SQUID potential}
U_J(\hat{\phi},\Phi_{\text{ext}}) = -E_J(\Phi_{\text{ext}})\cos\left(\hat{\phi}\right),
\end{equation}
where $E_J(\Phi_{\text{ext}}) = 2E_{J0}\left|\cos\left(\pi\Phi_{\text{ext}}/\Phi_0\right)\right|$ with $E_{J0}$ the Josephson energy of a single junction. For small $\hat{\phi}$, we only keep the cosine potential to fourth order and treat the quartic term as a perturbation. In combination, we obtain:
\begin{align}
H &\simeq 4E_C (\hat{n} - n_g(t))^2 + \frac{E_J(\Phi_{\text{ext}})}{2} \hat{\phi}^2 - \frac{E_J(\Phi_{\text{ext}})}{24} \hat{\phi}^4 - \xi I_d(t)\hat{\phi}.
\end{align}
We second quantize this Hamiltonian by writing
$
\hat{\phi} = \phi_{\text{zpf}} (\hat{a} + \hat{a}^\dagger)$ and $
n = i n_{\text{zpf}} (\hat{a}^\dagger - \hat{a}),
$
where $\phi_{\text{zpf}}$ and $n_{\text{zpf}}$ are $\left(\frac{2E_C}{E_J}\right)^{1/4}$ and $\frac{i}{2}\left(\frac{E_J}{2E_C}\right)^{1/4}$, respectively. $\hat{a}^\dagger$ and $\hat{a}$ are the raising and lowering
operators and satisfy the relation of $[\hat{a},\hat{a}^\dagger] = 1$. We ignore the constant term and the final Hamiltonian becomes:
\begin{equation}
H = \hbar\omega_q \hat{a}^\dagger \hat{a} -E_C \hat{a}^\dagger \hat{a}^\dagger \hat{a} \hat{a}- i 8E_C n_g(t) n_{\text{zpf}} (\hat{a}^\dagger - \hat{a}) - \xi I_d(t) \phi_{\text{zpf}} (\hat{a} + \hat{a}^\dagger),
\end{equation}
where $\hbar\omega_q = \sqrt{8E_C E_J(\Phi_\text{ext})}-E_C$.

We consider a qubit formed by the two lowest levels. The Hamiltonian in the qubit subspace can be approximated as:
\begin{equation}
H_{\text{qubit}} \simeq \hbar\omega_q \hat{\sigma}_z + 8E_C n_g(t) n_{\text{zpf}} \hat{\sigma}_y - \xi I_d(t) \phi_{\text{zpf}} \hat{\sigma}_x,
\end{equation}
where $\hat{\sigma}_i$ ($i=x,y,z$) are Pauli operators. With this qubit Hamiltonian, we can realize the following operations:

\begin{itemize}
\item \textbf{Single qubit gate}:

For a drive current $I_d(t) = I_0\cos(\omega_d t)$ or drive voltage $V_g(t) = V_0\cos(\omega_d t)$ resonant with the qubit, this term drives Rabi oscillations and can be used to perform arbitrary single-qubit gate with the Pauli operators.

\item \textbf{Qubit readout}:

The qubit is capacitively coupled to a readout resonator with a frequency of $\omega_r$. The combined Hamiltonian is:
\begin{equation}
H = H_{\text{qubit}} + \hbar\omega_r \hat{b}^\dagger \hat{b} + \hbar g (\hat{a}+ \hat{a}^\dagger)(\hat{b} + \hat{b}^\dagger),
\end{equation}
where $g$ is the capacitive coupling strength, and $\hat{b}$ ($\hat{b}^\dagger$) are the resonator annihilation (creation) operators. After the rotating wave approximation and in the dispersive regime ($|\Delta| = |\omega_q - \omega_r| \gg g$), a Schrieffer-Wolff transformation yields:
\begin{equation}
H_{\text{eff}} \simeq H_{\text{qubit}} + \hbar(\omega_r+\chi \hat{\sigma}_z) \hat{b}^\dagger \hat{b},
\end{equation}
with dispersive shift $\chi = g^2/\Delta$. The resonator frequency is qubit-state dependent. When the qubit is in the ground (excited) state, the resonator frequency is $\omega_r+\chi$ ($\omega_r-\chi$). Thus, we can directly measure the resonator frequency to detect the qubit state.


\item \textbf{Charge-parity detection}:

For slowly varying $n_g$ or $\Phi_\text{ext}$, we can tune the qubit frequency by modulating $E_C$ or $E_J$, as shown in Eq.~\eqref{eq:Hamiltonian} and Eq.~\eqref{eq:SQUID potential}. When a QP tunnel across the Josephson junction, it will change the charge of the capacitor pad. We define the charge parity of the qubit to be $P=\pm 1$ for even or odd charge parity. The Hamiltonian describing this qubit system becomes
\begin{equation}
    H=4E_C\left(\hat{n}-n_g(t)+\dfrac{P-1}{4}\right)^2-E_J(\Phi_\text{ext})\cos\hat{\phi}\,,
\end{equation}
where we ignore the current-drive and the qubit-readout terms for convenience. This coincides with the parity-dependent Hamiltonian in the main text.

The parity-dependent transition frequency of qubit is defined as $f_{\mathrm{o}}$ and $f_{\mathrm{e}}$ corresponding to the odd and even charge-parity branches, respectively, as shown in Fig. 1(e) in the main text. The frequency difference $2\Delta f=|f_{\mathrm{e}}-f_{\mathrm{o}}|$ can be exploited to map the charge-parity states onto the eigenstates of the qubit through a Ramsey-based sequence, as shown in Fig.~\ref{fig:setup}(b). The basic idea is to first use a single-qubit gate to rotate the qubit state onto the $X$ axis of the Bloch sphere. Owing to the parity-dependent frequency splitting $\Delta f$, after waiting for a time $1/(4\Delta f)$ the odd- and even-parity states have evolved to the $-Y$ and $+Y$ axes, respectively. A second rotation then maps these states onto the $Z$ axis, where a projective measurement in the computational basis yields an outcome $\ket{0}$ or $\ket{1}$, from which the charge parity can be inferred. The qubit is subsequently reset to $\ket{0}$, and the sequence is repeated to realize continuous monitoring of the charge parity.

As illustrated in Fig.~\ref{fig:setup}(c), the operation of the quantum parity detector (QPD) based on a qubit array is as follows. The charge parity of all qubits is monitored simultaneously. In the absence of a signal, only occasional parity switches occur on a few qubits due to residual QP tunneling, and the overall parity pattern remains stable. When a signal arrives, the charge parity of many qubits switches within a short time, leading to a instable pattern. As the QP burst generated by the signal relaxes back to the background level, the parity pattern also becomes stable. By counting, for each qubit, the number of QP tunneling events within the QP relaxation time, one can identify the occurrence of a signal. Because the signal produces correlated responses across multiple qubits, combining the information from the entire array further enhances the signal-to-noise ratio and yields a schematic response curve. By varying the input phonon energy and measuring the corresponding response, the detector can be calibrated.

\item \textbf{Gate control}:

In addition, experiments have shown that the offset charge drifts (with a typical interval of order tens of minutes~\cite{Christensen2019Anomalous,wilen2021correlated}) will change the $\Delta f$, so that the detector must be recalibrated and may even fail to operate when $\Delta f = 0$. To overcome this issue, we utilize the additional gate line of each qubit, which enables us both monitoring and control of the offset charge $n_g(t)$ in a short time (down to below $\sim 10~\mathrm{s}$). This design allows the detector to operate continuously for weeks to months, requiring only a few very short recalibration steps.
\end{itemize}

\section{Parameters for detector design and simulations}

We list the values of the parameters for the detector design, simulations of phonon transport and calculations of the QP tunneling in Table~\ref{tab:parameters}.

\begin{table}[!h]
\setlength{\tabcolsep}{20pt}
\renewcommand{\arraystretch}{1.3}
    \centering
    \begin{tabular}{c|c}
        {\bf Parameter} & {\bf Value} \\ \hline
        Size of thin substrate (length$\times$width$\times$height [mm]) & $21.2\times 22.6\times 0.43$\\ \hline
        Size of thick substrate (length$\times$width$\times$height [mm]) & $21.2\times 22.6\times 20$\\ \hline       
        Number of qubits in the chip & 96 \\ \hline
        Number of In supports & 4\\ \hline
        Area of Ta absorber film in a qubit [$\mu$m$^2$] & 87394\\ \hline
        Area of an In support [$\mu$m$^2$] & 946266\\ \hline
        Height of an In support [$\mu$m] & 10\\ \hline
        Thickness of Ta film [nm] & 200  \\ \hline
        Volume of Al trap [$\mu{\rm m}^3$] & 10\\ \hline
        Phonon decay constant in substrate $A$ [s$^4$] & $1.27\times 10^{-55}$~\cite{Kelsey:2023eax} \\ \hline
        Phonon scattering constant in substrate $B$ [s$^3$]& $2.5\times 10^{-44}$~\cite{wigmore2002scattering} \\ \hline
        Phonon absorption probability at the sapphire-Ta interface & 0.94 [Sec.~\ref{sec:phononinterface}]\\ \hline
        Phonon absorption probability at the sapphire-In interface & 0.86 [Sec.~\ref{sec:phononinterface}]\\ \hline        
        Maximum number of phonon reflections & 1000~\cite{CDMSsimulation} \\ \hline
        Partition of longitudinal phonon state & 0.137~\cite{Kelsey:2023eax} \\ \hline
        Partition of transverse fast phonon state & 0.351~\cite{Kelsey:2023eax}\\ \hline
        Partition of transverse slow phonon state & 0.511~\cite{Kelsey:2023eax}\\ \hline
        Al density of states $D_f$ [m$^{-3}\cdot$eV$^{-1}$] & $1.46\times 10^{28}$\\ \hline
        Operating temperature [mK] & 10 \\ \hline
        Superconducting gap of Ta [$\mu$eV] & 680 \\ \hline
        Superconducting gap of Al [$\mu$eV] & 190 \\ \hline
        Reference Josephson energy $E_J/h$ [GHz] & 6.14\\ \hline
        Pair breaking efficiency & 0.6~\cite{guruswamy2014quasiparticle,kozorezov2000quasiparticle}\\ \hline
        Trapping efficiency & 0.7 [Sec.~\ref{sec:trapefficiency}]\\ \hline
        Quasiparticle recombination time in Al [ms]~\cite{barends2009enhancement,wang2014measurement} & 1\\ \hline
         Time window $t_{\rm opr}$ [ms] & 1.5 \\ \hline        
        Fano factor $F$& 0.2~\cite{verhoeve2002superconducting} \\ \hline
        Detection fidelity $\mathcal{F}$ & 0.95~\cite{Wang:2024rjw,marxer2025above}
    \end{tabular}
    \caption{Parameters used in this work. The references are included where applicable.}
    \label{tab:parameters}
\end{table}

\section{Phonon absorption probability at the interface}
\label{sec:phononinterface}

After propagation in the substrate, the phonons will arrive at the interface between the substrate and the superconducting Ta film or the In pillar. The probability for the phonon to be absorbed at the interface $P_{\rm abs}$ depends on two factors, the transmission probability $T_{\rm trans}$ and the probability to escape back to the substrate $P_{\rm esc}$, with the relation
\begin{equation}
    P_{\rm abs}=T_{\rm trans}(1-P_{\rm esc})\,.
    \label{eq:absprob}
\end{equation}

The transmission probability of phonons from sapphire (Al$_2$O$_3$) into Ta and In can be estimated using the Acoustic Mismatch Model (AMM) \cite{Swartz1989rmp}. According to this model, the transmission coefficient $T_{\rm trans}$ at normal incidence is given by
\begin{equation}
    T_{\rm trans} = \frac{4Z_1Z_2}{(Z_1 + Z_2)^2}\,,
\end{equation}
where $Z_1$ and $Z_2$ are the acoustic impedance of sapphire and Ta (In), respectively, calculated $Z = \rho v_s$, where $\rho$ is the density of the material and $v_s$ is the speed of sound in the material. Given the material parameters (using longitudinal speed of sound) $\rho_{\text{Al}_2\text{O}_3} = 3.95\times 10^3\,\mathrm{kg/m^3}$, $v_{s,\,\text{Al}_2\text{O}_3} = 1.04\times 10^4\,\mathrm{m/s}$ \cite{every1984ballistic}, $\rho_{\text{Ta}} = 16.6\times 10^3\,\mathrm{kg/m^3}$, and $v_{s,\,\text{Ta}} = 4.1\times 10^3\,\mathrm{m/s}$~\cite{Orlikowski2006prb}, the resulting phonon transmission probability at the sapphire--Ta interface is calculated to be $T_{\rm trans} = 0.94$. For phonons at $\mathcal{O}(\mathrm{meV})$ energy and an interface with a surface roughness of less than $0.5\,\mathrm{nm}$ over a $5\times5\,\mu\mathrm{m}^2$ area (as specified by the substrate supplier), it is expected that the actual probability of phonon transmission closely approaches this theoretical value \cite{Wen2009prl, Gelda2018prb}. Similarly, the transmission probability at the sapphire--In interface is 0.86 using $\rho_{\text{In}} = 7.31\times 10^3\,\mathrm{kg/m^3}$ ~\cite{samsonov2012handbook} and the sound speed $v_{s,\,\text{In}} = 2.57\times 10^3\,\mathrm{m/s}$ computed following the parameters in~\cite{application_note}.

After entering the film, the phonon has a probability to escape back to the substrate, which can be computed by comparing the mean free path of phonon in the film and the thickness $d$ of the film. The escape probability at normal incidence is therefore~\cite{Linehan:2025suv}
\begin{equation}
    P_{\rm esc}=\exp\left(-\dfrac{2d E_{\rm ph}}{\pi v_s\tau_0\Delta}\right)\,,
\end{equation}
where $\tau_0$ is the phonon scattering time.
For the Ta film $d=200\,\text{nm}$ in our design and $\tau_0=2.3\times10^{-11}\,\text{s}$ \cite{Kaplan1976prb}, $P_{\rm esc}\simeq 0$ for a few meV phonon. Similar behavior is expected for the In pillars, since the value of $d$ is much larger. In combination, we find $P_{\rm abs,\,Ta}=0.94$ and $P_{\rm abs,\,In}=0.86$ using Eq.~\eqref{eq:absprob}.

The validity of our treatmeat can also be verified from the literature.
It can be confirmed that Eq.~\eqref{eq:absprob} yields the absorption probability $P_{\rm abs}=0.3$--$0.53$ for 4--10~meV phonon with a 60~nm Al film at Si--Al interface, consistent with the experimental measurements in~\cite{Martinez:2018ezx}. The is lower than the absorption probability at the sapphire--Ta interface in this work, mainly due to the fact that low thickness of the Al film there increases the probability for phonons to escape back to the substrate.

\section{Phonon transport and quasiparticle number density}

\begin{figure}[!htb]
    \centering
    \includegraphics[width=0.5\columnwidth]{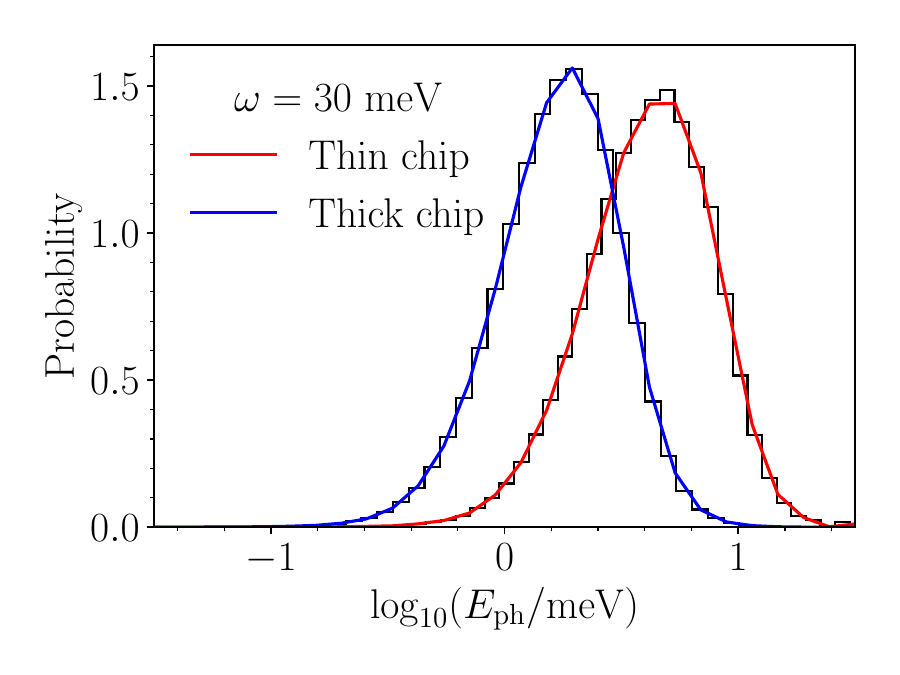}   
    \caption{The probability distribution of the energy of phonons entering the Ta films from phonon propagation simulations. We simulate phonons uniformly located in the sapphire substrate with random polarization states. The energy of the primary phonon is 30~meV. The histogram show the binned probability and the lines are fits using a Gaussian kernel.}
    \label{fig:energydist}
\end{figure}

We simulate the propagation of phonons with energy $\omega$ uniformly injected in the substrate as described in the main text, and collect secondary phonons that enter Ta films of the qubits.  The energy distribution of individual phonons $E_{\rm ph}$ in the films is depicted in Fig.~\ref{fig:energydist}. Although we show the statistics of phonons in all of the qubits, the distribution is not very different across different qubits as the location of the primary phonons is randomized in the substrate.

 We find the average phonon energy in the films is about 4.9 (2.0) meV in the thin-chip (thick-chip) detector. This is not unexpected as high-energy phonons will quickly downconvert to lower energy phonons in propagation at a rate $\Gamma=A\nu^5$. The mean free path for longitudinal phonon decay in sapphire increases from 0.11 mm at $E_{\rm ph}=10$~meV, to 333~mm at $E_{\rm ph}=2$~meV, which is much larger than the size of the substrate adopted in this work. High-energy transverse phonons are also likely to change modes at a rate $\Gamma=B\nu^4$ when scattering with isotopes. Once the mode switches to longitudinal, they may also decay in propagation.

The QPs created by phonons will recombine and go through interconversion  with phonons and finally dissipate away. The process takes place at a characteristic time scale $\tau_{\rm qp}\sim 1$~ms for Al~\cite{barends2009enhancement,wang2014measurement}. The evolution of the QP number density is then
\begin{equation}
    n_{\rm qp}(t)=n_{\rm qp}e^{-t/\tau_{\rm qp}}\,,
\end{equation}
where $n_{\rm qp}$ is determined by the phonons absorbed in a qubit as described in the main text.

\section{Trapping efficiency}
\label{sec:trapefficiency}

We describe the QP dynamics in the Ta--Al structure using the following two-dimensional diffusion equation:
\begin{equation}
\frac{\partial n(\bm{r},t)}{\partial t} = D_0 \nabla^2 n(\bm{r},t) - \frac{n(\bm{r},t)}{\tau(r^2)} + s(\bm{r},t),
\end{equation}
where $n(\bm{r},t)$ is the QP density profile, and the first term on the right-hand side represents diffusion, with $D_0$ denoting the Ta diffusion coefficient. The second term accounts for two distinct processes: in the Ta pad region, $\tau(r^2)\equiv\tau_r$ is the characteristic time for QP recombination into Copper pairs within Ta; in the Al trap, $\tau(r^2)\equiv\tau_s$ is the characteristic time for QP relaxation below the Ta superconducting gap through emitting a phonon. Thus in the Al region, the term $n(\bm{r},t)/\tau_s$ represents the QP-trapping rate per unit area, which, after integration over time and space, yields the total number of QPs trapped in the Al region. The last term, $s(\bm{r},t)$, denotes the time-dependent QP source, assumed to be a finite pulse such that the QP profile $n(\bm{r},t)$ remains unchanged in the distant past and future. The time-integrated QP dynamics equation then becomes
\begin{equation}
0 = D_0 \nabla^2 \phi(\bm{r}) - \frac{\phi(\bm{r})}{\tau(r^2)} - G(\bm{r}),
\end{equation}
where $\phi(\bm{r})=\int n(\bm{r},t) \mathrm{d}t$ is the time-integrated QP density and $G(\bm{r})=\int s(\bm{r},t) \mathrm{d}t$ is the spatially dependent source term. 

\begin{figure}[!htb]
    \centering

    \includegraphics[width=0.6\columnwidth]{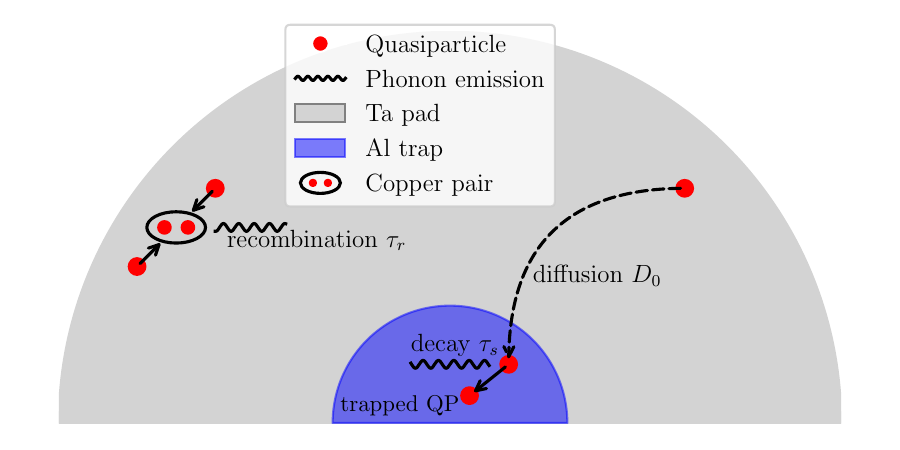}

    \caption{ QP dynamics considered including diffusion and recombination in Ta and trapping in Al. }
    \label{fig:QPdynamics}
\end{figure}

For simplicity, we consider a semicircular Ta pad of radius $l$ and Al trap of radius $a$ with reflective boundary conditions $\hat{n}\cdot\nabla\phi = 0$, as illustrated in Fig.~\ref{fig:QPdynamics}. The time-integrated source term $G(\bm{r})$ is modeled as a Gaussian wave packet centered at $(x_0,y_0)$:
\begin{equation}
G(\bm{r}) = \frac{N}{2\pi\sigma^2} \exp\!\left[ -\frac{(x - x_0)^2 + (y - y_0)^2}{2\sigma^2} \right],
\end{equation}
where $N$ is the total number of generated QPs, and the Gaussian half-width is chosen as $\sigma = 10^{-2} \,l$. Recall that $\phi(\bm{r})=\int n(\bm{r},t) \mathrm{d}t$ and $n(\bm{r},t)/\tau_s$ represents the QP-trapping rate, the trapping efficiency $\eta_{\rm trap}$ is then given by the normalized total $\phi(\bm{r})$ within the aluminum trap:
\begin{equation}
\eta_{\rm trap} = \frac{1}{\tau_s N} \int_0^\pi \int_0^a \phi(r\cos\theta, r\sin\theta)\, r\, \mathrm{d}r\, \mathrm{d}\theta,
\end{equation}
which quantifies the fraction of QPs trapped in the Al region relative to the total number of generated QPs $N$.

\begin{table}[!h]
\setlength{\tabcolsep}{20pt}
\renewcommand{\arraystretch}{1.3}
\centering
\begin{tabular}{c|c|c}
    {\bf Parameter name} & {\bf Meaning} & {\bf Value} \\ \hline
    $ l $         & Ta pad radius [$\mu\text{m}$]                 & 166.788 \\ \hline
    $ a $         & Al trap radius [$\mu\text{m}$]           & $l/41.8$ \\ \hline
    $ D_0 $       & Diffusion coefficient [$\mu\text{m}^2\cdot\mu\text{s}^{-1}$]         & $1.5\times10^3\text{--}3.0\times10^3$~\cite{Friedrich:1997,Lissitski2004}\\ \hline
    $ \tau_r $    & Recombination lifetime in tantalum absorber [$\mu\text{s}$] & $83$~\cite{Li2003jap} \\ \hline
    $ \tau_s $    & Decay lifetime in aluminum trap [$\mu\text{s}$]      & $0.01$~\eqref{eq:taus}\\ \hline
    $ \sigma $    & Gaussian half-width            & $ 10^{-2} \,l$ \\ \hline
    $ (x_0, y_0) $& Center of Gaussian packet      & uniformly sampled \\ 
\end{tabular}
\caption{
Physical parameters used in the model. The choice of $l$ and $a$ corresponds to the area of Ta absorber film in a qubit and volume of Al trap in Table~\ref{tab:parameters}. The diffusion coefficient  $D_0$ is sensitive to the fabrication quality of the metallic film. Ref.~\cite{den2000} reports $1.5$--$2.0\times10^{3}\ \mu\mathrm{m}^2/\mu\mathrm{s}$. We note that the diffusion constant is proportional to the residual-resistance-ratio $\mathrm{RRR}\equiv\rho(\text{300 K})/\rho(\text{4.5 K})$ of the film~\cite{den2000}, and in~\cite{den2000} $\mathrm{RRR}=45$. Ref.~\cite{Lissitski2004} reports a Ta film with RRR = 75 with an improved preparation process, corresponding to $D_0=2.5$--$3.3\times10^{3}\ \mu\mathrm{m}^2/\mu\mathrm{s}$.}
\label{tab:physicalparams}
\end{table}

Following \cite{Kaplan1976prb}, at zero temperature, the decay rate of a QP with energy $\omega$ in a BCS superconductor with gap $\Delta$ is obtained by integrating over the energy of the emitted phonon $\Omega$:

\begin{equation}
    \tau_s^{-1}(\omega) = \tau_0^{-1} \left(\frac{\Delta}{k_B T_c}\right)^3  \int_{0}^{\omega - \Delta} \frac{ \mathrm{d}\Omega \,  \Omega^2}{\Delta^3} \,  \frac{\omega - \Omega}{\sqrt{(\omega - \Omega)^2 - \Delta^2}}  \left( 1 - \frac{\Delta^2}{\omega \, (\omega - \Omega)} \right)\,, 
    \label{eq:taus}
\end{equation}
where for a BCS superconductor at zero temperature, the superconducting gap satisfies $\Delta = 1.76\,k_B T_c$.
For $\omega = 680\,\mathrm{\mu eV}$, $\Delta = 190\,\mathrm{\mu eV}$, and $\tau_0 \simeq 438\,\mathrm{ns}$ (corresponding to the superconducting gaps of Ta and Al and the characteristic scattering time of Al), we obtain $\tau_s \simeq 10\,\mathrm{ns}$.

In the limit $D_0 \rightarrow \infty$, corresponding to ultra-fast diffusion, the QP density profile $n(\bm{r},t)$ becomes uniform instantaneously. In this case, the trapping efficiency depends only on the relative size of the trap $a^2/l^2$ and the ratio $\tau_s/\tau_r$,
\begin{equation}
\eta_{0} = \frac{a^2/\tau_s}{a^2/\tau_s + (l^2 - a^2)/\tau_r}.
\label{eq:eta0}
\end{equation}

\begin{figure}[!htb]
    \centering
    \includegraphics[width=0.49\columnwidth]{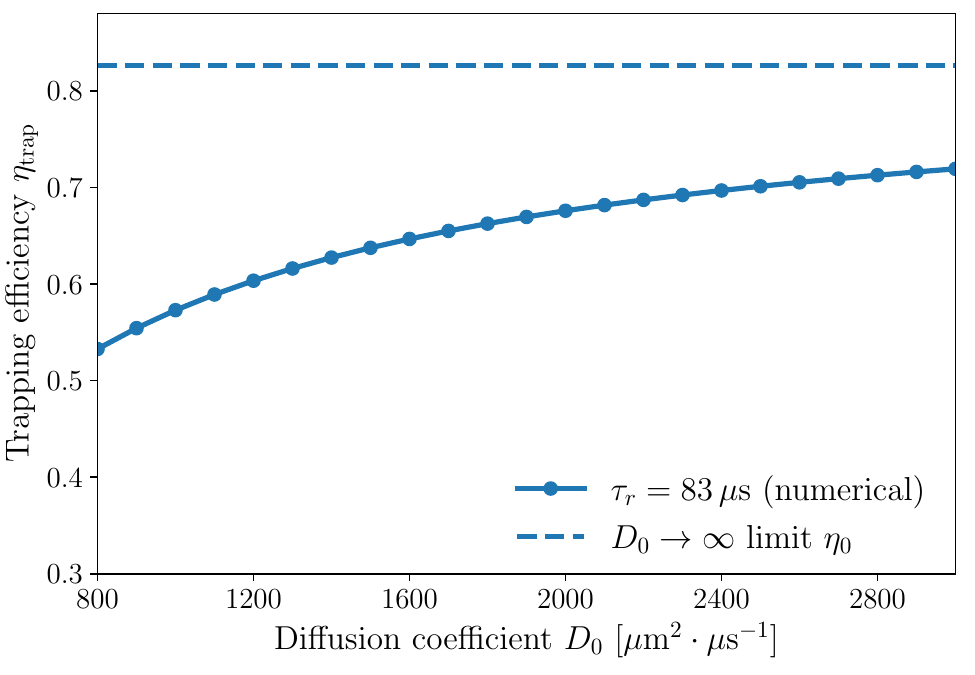}    
    \caption{Using physical parameters specified in Table~\ref{tab:physicalparams}, we sample source points $(x_0,y_0)$ uniformly over the Ta pad area to obtain the spatially-averaged trapping efficiency $\eta$ shown above. To account for the uncertainties in $D_0$, we vary its value in between $1.5\times10^3$ and $3.0\times10^3\,\mu\mathrm{m}^2\,\mu\mathrm{s}^{-1}$.}
    \label{fig:trapeff}
\end{figure}

We use the parameters in Table \ref{tab:physicalparams} to give estimations on trapping efficiency, see results shown in Fig.~\ref{fig:trapeff}. For reasonable parameters $D_0=2.5\times10^3\,\mu\text{m}^2\cdot\mu\text{s}^{-1}  $ and $\tau_r=83\,\mu\text{s}$, we can achieve $\eta_{\rm trap}\simeq 0.70$ as claimed in the main text.

If we conservatively choose the parameters reported in~\cite{Li2003jap}, i.e., $\tau_r=83\ \mu\mathrm{s}$ and $D_0=820\ \mu\mathrm{m}^2/\mu\mathrm{s}$, the trapping efficiency is evaluated to be 0.54. As the number of tunnelings is proportional to $\eta_{\rm trap}$, the lower $\eta_{\rm trap}$ will lead to slightly lower efficiencies below 30~meV. However, as we will discuss in Sec.~\ref{sec:n0effect}, this lies in the acoustic phonon band, which is kinematically inaccessible to light dark matter, and has little effect on the sensitivity.

\section{Quasiparticle tunneling rate and energy reconstruction}
\label{sec:tunnelingrate}

The QPs injected from the Ta film will rapidly reach thermal equilibrium in the Al trap, but not chemical equilibrium. The phonon-initiated QPs induce a shift in the chemical potential $\delta\mu$ to the Fermi-Dirac distribution~\cite{owen1972superconducting}
\begin{equation}
    f_{\rm FD}(E_{\rm qp}-\delta\mu)=\dfrac{1}{1+\exp[{(E_{\rm qp}-\delta\mu)/k_B T]}}\,.
\end{equation}
By relating the distribution to the number density of QPs we can solve for the chemical potential, which gives~\cite{palmer2007steady}
\begin{equation}
    \delta\mu\simeq \Delta_{\rm trap}+k_BT\ln(n_{\rm qp
    }/\mathcal{N}_{\rm qp})\,,
    \label{eq:deltamu}
\end{equation}
in the low temperature limit $T\ll \Delta_{\rm trap}$, where $\mathcal{N}_{\rm qp}=\sqrt{2\pi\Delta_{\rm trap}k_B T}D_f$ represents the number of QP states available in the trap, and $D_f=3n_e/2E_f$ is the normal metal density of states in the trap, with $n_e$ the number density of electrons near the Fermi surface and $E_f$ the corresponding Fermi energy.

The QP tunneling rate can be computed from Fermi's Golden Rule, by integrating the distribution of QPs. Assuming that the QPs tunnel from trap 1 to trap 2 on two sides of the Josephson junction, the tunneling rate is~\cite{shaw2008kinetics,shaw2009quantum,lutchyn2007kinetics}
\begin{equation}
    \Gamma_{\rm tun}=\dfrac{16E_J}{h\Delta_{\rm trap}}\int\limits_\Delta^\infty \mathrm{d}E\dfrac{EE'-\Delta_{\rm trap}^2}{\sqrt{(E^2-\Delta_{\rm trap}^2})(E'^2-\Delta_{\rm trap}^2)}
    \times f_{\rm FD}(E-\delta\mu_1)[1-f_{\rm FD}(E'-\delta\mu_2)]\,,
    \label{eq:tunnelingratefull}
\end{equation}
where $E'=E+\delta$, and $\delta$ is the energy shift before and after the tunneling. $E_J=h\Delta_{\rm trap}/8R_N e^2$ is the Josephson energy and $R_N$ is the normal state resistance of the junction. The square bracket takes into account the Pauli blocking effect.  In the low temperature limit the QP occupation is quite low and $f_{\rm FD}(E'-\delta\mu_2)\simeq 0$ is a good approximation. Plugging Eq.~\eqref{eq:deltamu} into Eq.~\eqref{eq:tunnelingratefull} we arrive at the tunneling rate in the main text
\begin{equation}
    \Gamma_{\rm tun}(t)\simeq \dfrac{16E_J k_B T n_{\rm qp}(t)}{\mathcal{N}_{\rm qp}h\Delta_{\rm trap}}\,,
\end{equation}
where $n_{\rm qp}(t)=n_{\rm qp}e^{-t/\tau_{\rm qp}}$ decays exponentially due to QP recombination. Note that $T$ should be interpreted as the effective QP temperature, which could be substantially higher than the mixing chamber stage of the dilution refrigerator, due to imperfect shielding and filtering, and other QPs not in chemical equilibrium~\cite{lvov2025thermometry}. Measurements suggest the tunneling rates show an upturn at around 100~mK, above which the thermal-induced QPs dominate over the nonequilibrium QPs~\cite{kurter2022quasiparticle,Serniak2018hot}. We therefore take $T=100$~mK. As both the signal- and residual-QP-induced tunneling rates scale with $\sqrt{T}$, a different choice of $T$ will not change the signal-to-background ratio. The effective QP temperature will also be calibrated in the experiment.

In one measurement, we count the number of tunneling signals in an operation time window $t_{\rm opr}$, i.e.
\begin{equation}
    N_s=\int_0^{t_{\rm opr}} \Gamma_{\rm tun}(t)\mathrm{d}t 
    =\dfrac{16E_J k_B T n_{\rm qp}}{\mathcal{N}_{\rm qp}h\Delta_{\rm trap}}t_{\rm eff}\,,
    \label{eq:Ntun}
\end{equation}
where the effective measurement time $t_{\rm eff}=\int n_{\rm qp}(t)/n_{\rm qp}dt$. As a result of finite $\tau_{\rm qp}$, we have $t_{\rm eff}<t_{\rm opr}$. In simulations, we draw the number of tunnelings from Poisson distribution both for signals and the residual QP background.

\begin{figure}[!htb]
    \centering
    \includegraphics[width=0.49\columnwidth]{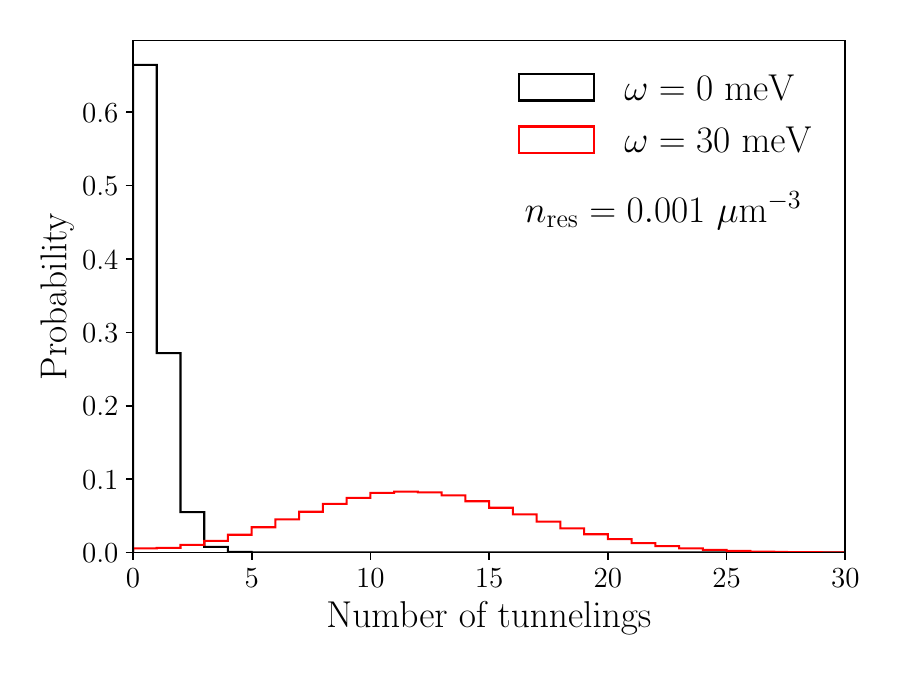}
    \includegraphics[width=0.49\columnwidth]{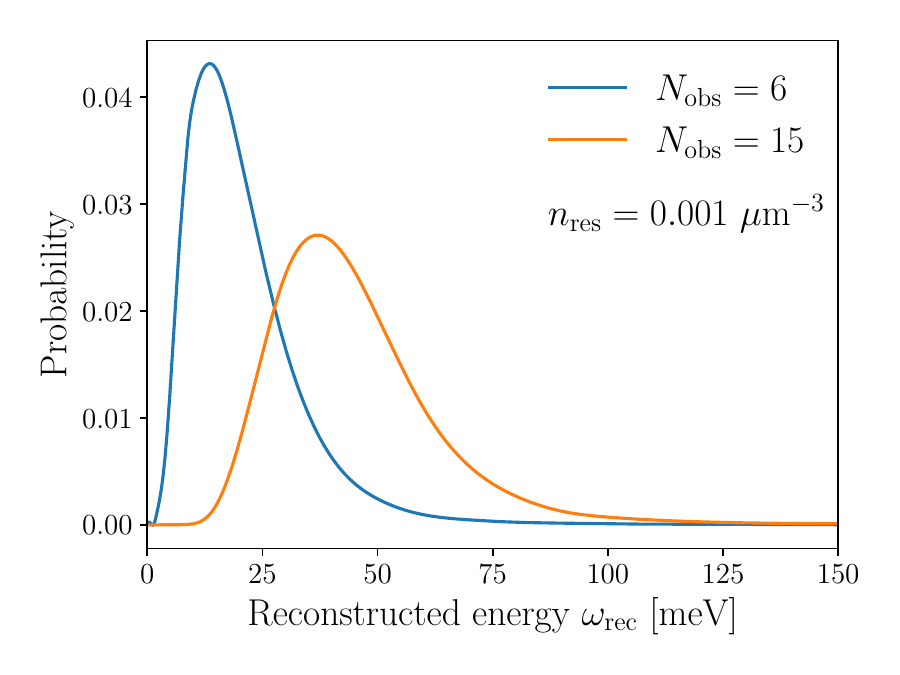}
    \caption{{\it Left:} The probability distribution of the total number of tunnelings for the primary phonon energy $\omega=0$~meV (black) and $\omega=30$~meV (red) in the substrate in the measurement time window with a thin chip. We assume the total QP efficiency $\eta\equiv \eta_{\rm pb}\eta_{\rm trap}=0.42$ and the residual QP density $n_{\rm qp}=0.001~\mu\mathrm{m}^{-3}$. {\it Right:} The probability distribution of the reconstructed energy for a total number of 6 (blue) and 15 (orange) QP tunnelings in all qubits.}
    \label{fig:Ntundist}
\end{figure}

We show the distribution of the total number of tunnelings (signal+background) in all the qubits in the measurement time window from simulations in the left panel of Fig.~\ref{fig:Ntundist}. For $\omega=0$, i.e., no phonon energy deposition in the substrate, the probability distribution is strongly peaked near 0, while nonzero tunnelings come from the contribution of the residual QP density. For $\omega=30$~meV the distribution is nearly Gaussian and is phonon-initiated tunnelings dominate. The average number of tunnelings is much larger than that for $\omega=0$, ensuring good signal-background discrimination and high detection efficiency at $\omega=30$~meV.

The distribution also allows us to construct the probability of the reconstructed energy $\omega_{\rm rec}$, see the right panel of Fig.~\ref{fig:Ntundist}. Given the total number of tunnelings $N_{\rm obs}$ in a measurement, we find $\mathcal{L}(N_{\rm obs}|\omega)$ from simulations. $\omega_{\rm rec}$ is obtained by maximizing the likelihood, and the energy resolution is computed as the full width at half-maximum (FWHM) of the probability.

\section{Impact of different residual quasiparticle densities and detector design}
\label{sec:n0effect}

\begin{figure}[!htb]
    \centering
    \includegraphics[width=0.32\columnwidth]{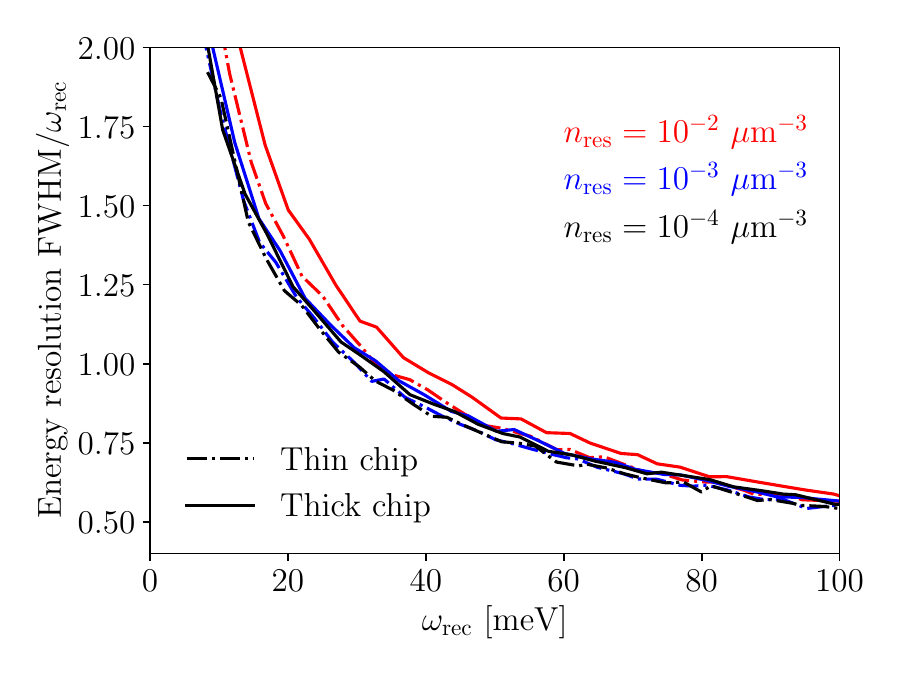}
    \includegraphics[width=0.32\columnwidth]{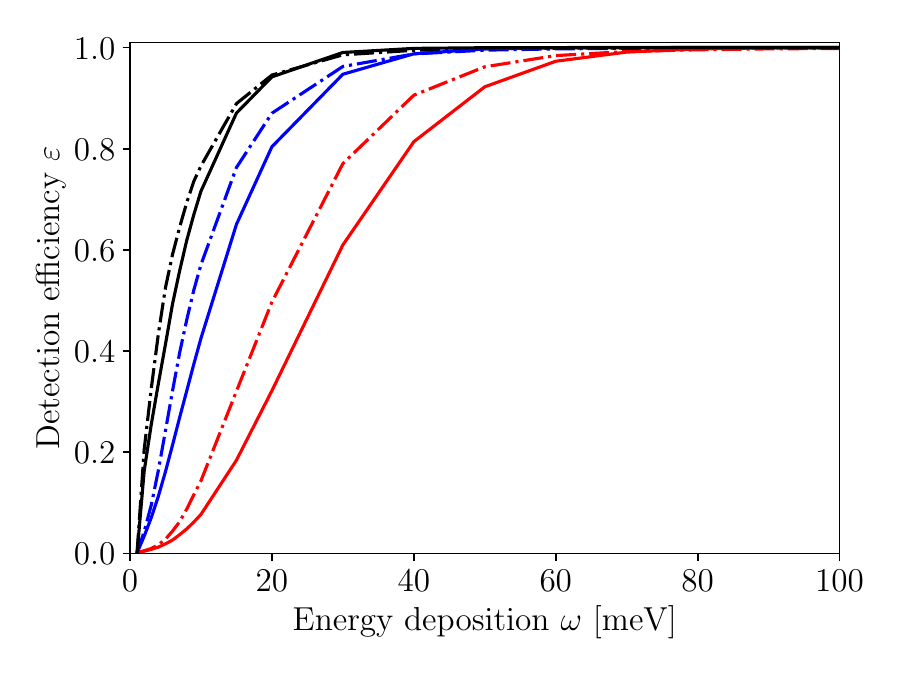}
    \includegraphics[width=0.32\columnwidth]{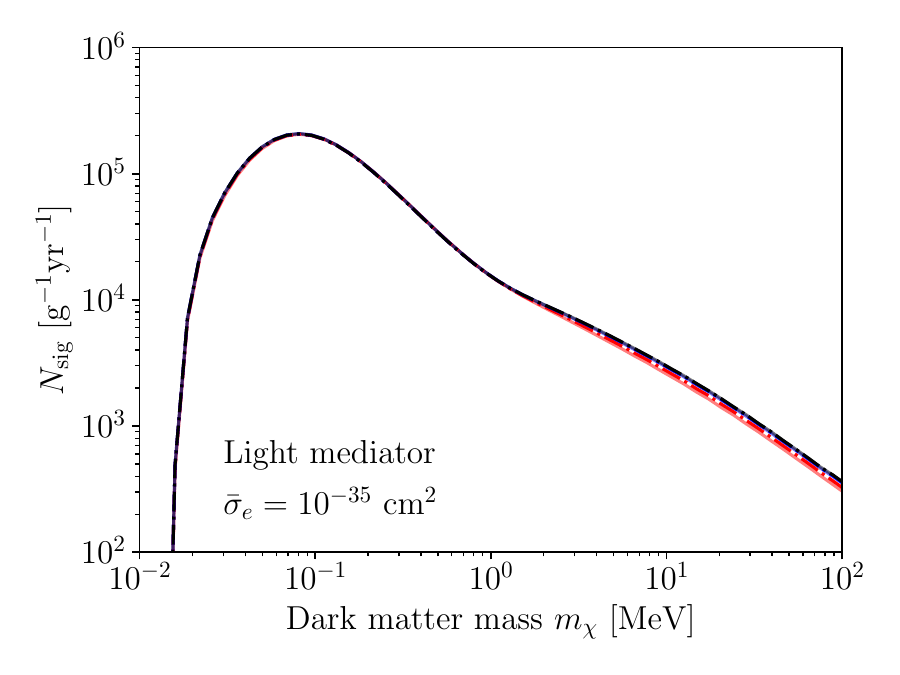}
    \caption{{\it Left:} The energy resolution as a function of the reconstructed phonon energy for a thin-chip (dash-dotted line) and thick-chip (solid line) detector.  The red, blue and black lines correspond to the residual QP density of $10^{-2}~\mu\mathrm{m}^{-3}$, $10^{-3}~\mu\mathrm{m}^{-3}$, and $10^{-4}~\mu\mathrm{m}^{-3}$, respectively. {\it Middle:} The detection efficiency as a function of the energy deposition. The color coding is the same as the left panel. {\it Right:} The number of signal events as a function of DM mass with the thin-chip setup assuming a light mediator. We assume the reference DM-electron scattering cross section $\bar{\sigma}_e=10^{-35}~{\rm cm}^{2}$. The color coding is the same as the left panel.}
    \label{fig:resolution}
\end{figure}

We now examine the impact of the residual QP number density, $n_{\rm res}$. As discussed in the main text and Sec.~\ref{sec:tunnelingrate}, the background tunneling rate---and consequently the number of background tunnelings during a single measurement---is proportional to $n_{\rm res}$. The left panel of Fig.~\ref{fig:resolution} shows the energy resolution for different values of $n_{\rm res}$. As expected, a larger residual density degrades both the energy reconstruction and the resulting resolution. A higher background tunneling rate also implies that a larger energy deposition is required to be statistically distinguished from the background. Consequently, the detection efficiency decreases with increasing $n_{\rm res}$, as illustrated in the middle panel of Fig.~\ref{fig:resolution}.

The influence of detector geometry can also be inferred from Fig.~\ref{fig:resolution}. In a thicker substrate, phonons experience more diffusion and decay than in a thinner one, leading to poorer energy resolution and slightly reduced detection efficiency. However, the change in detection efficiency is of order $\mathcal{O}(1)$ and does not significantly affect the overall sensitivity, apart from the difference in exposure.

We can also observe this trend from the number of signal events. In the analysis, we restrict the energy deposition in the range where the detection efficiency satisfies $\varepsilon(\omega) \gtrsim 0.1$, and consider reconstructed energies between 2 and 100~meV. Increasing either the residual QP density or the chip size reduces the signal event rate by up to a factor of a few, primarily due to the corresponding decrease in detection efficiency.

Note that energy depositions $\omega \lesssim 30~\mathrm{meV}$ contribute little to the total event rate for the sapphire target, as this range lies in the acoustic phonon regime, which is typically kinematically inaccessible to light DM with small momentum transfer. In contrast, as illustrated in Fig.~\ref{fig:phononbands}, optical phonons originate at zero momentum transfer and dominate the scattering rate. Consequently, lowering the detection threshold below 30~meV does not significantly improve the sensitivity for sapphire. However, substrates with lower optical phonon energies could experience a greater enhancement in sensitivity with reduced thresholds, and hence are more suitable for detecting even lighter DM with smaller energy deposition.

\section{Discussions on potential background}

We have included the background from infrared photons which contributes to the residual QP density, and the scattering of high energy photons with phonon final states, when calculating the sensitivity in this work. The readout noise is accounted for in the detection fidelity. We list all possible background sources below. Note that the background level in real experiments may be substantially different from the background rate adopted in this work, and the sensitivity will be scaled correspondingly.

\begin{itemize}
    \item {\it Cosmic rays and radioactive decay.} 
    Cosmic rays and their secondary particles can ionize the substrate and produce electron--hole pairs, which subsequently generate cascades of phonons. These phonons disrupt the quantum coherence of superconducting qubits by inducing QP bursts, leading to correlated errors in quantum computers~\cite{vepsalainen2020impact,wilen2021correlated,martinis2021saving,mcewen2022resolving,Li:2024dpf,harrington2025synchronous}. Similarly, $\gamma$ rays from environmental radioactive decay can produce correlated errors as well~\cite{loer2024abatement}.

    Such backgrounds can be mitigated by applying lead shielding to suppress environmental $\gamma$ radiation and by operating the detector underground to reduce cosmic-ray exposure~\cite{bratrud2024first,cardani2021reducing,loer2024abatement,gordon2022environmental}. The experiment will be conducted at the China Jinping Underground Laboratory (CJPL), where the cosmic-ray and $\gamma$ backgrounds are expected to be extremely low~\cite{Ma:2020rpd}. Additional radioactivity from the dilution refrigerator or detector materials can be minimized through careful selection of low-radioactivity components~\cite{loer2024abatement}.

    It is worth noting that cosmic rays and radioactive decays typically deposit energies well above the keV scale---far exceeding the $\lesssim 100$~meV signal region of interest considered in this work.
    \item {\it Soft scattering of high energy photons.} As discussed above, high-energy photons with energies $\gtrsim$~keV are produced in radioactive decay processes. If such decays occur within the dilution refrigerator or the surrounding metallic enclosure, these photons can penetrate and scatter in the substrate~\cite{Berghaus:2021wrp}. Although most of the deposited energy is large, a small fraction of scatterings can transfer sub-eV recoil energies with low momentum exchange. These processes include photon--electron Rayleigh scattering, nuclear Thomson scattering, Delbrück scattering, and nuclear resonance scattering, with Rayleigh scattering dominating in this regime~\cite{Robinson:2016imi}.

    The resulting phonon excitations in the $\lesssim 100$~meV range are experimentally indistinguishable from the signal expected from light DM scattering or absorption. To model this background, we adopt the calculation of Ref.~\cite{Berghaus:2021wrp}, which uses the benchmark photon spectrum measured by the EDELWEISS Collaboration~\cite{EDELWEISS:2018tde} (see Fig.~\ref{fig:photonbackground}). The corresponding background rate is estimated to be $5.27~\mathrm{kg^{-1}\cdot yr^{-1}}$. We expect similar level of shielding to be installed in the experiment. An active veto could further suppress this background by rejecting events coincident with gamma detections.
\begin{figure}[!htb]
    \centering
    \includegraphics[width=0.45\columnwidth]{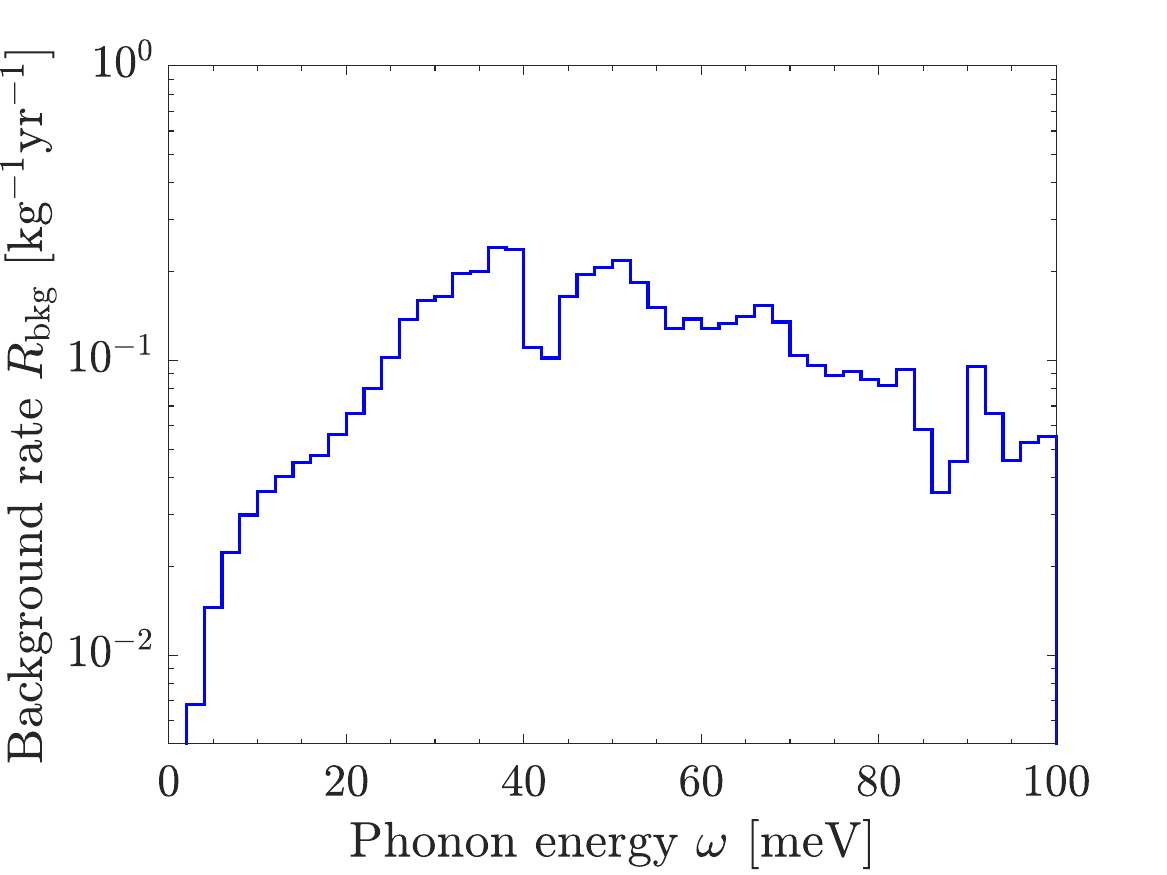}
    \caption{The phonon background rate from high energy photon scattering in the sapphire substrate, extracted from~\cite{Berghaus:2021wrp}.}
    \label{fig:photonbackground}
\end{figure}    
    \item {\it Infrared leakage.} Photons with energies near the pair-breaking threshold, i.e., $2\Delta_{\rm abs}$ or $2\Delta_{\rm trap}$, can efficiently break Cooper pairs and increase the residual QP density $n_{\rm res}$. Such radiation primarily originates from the higher-temperature stages of the dilution refrigerator and is transmitted through the coaxial wiring~\cite{Liu:2022aez}. This background can be mitigated by isolating and shielding the detector, coating the metallic housing with infrared-absorbing materials, and installing infrared filters along the wiring~\cite{Barends:2011uku,pan2022engineering}. 
    \item {\it Thermal stress.} An excess of low-energy (10--100 eV) events has been observed in several cryogenic DM detection experiments~\cite{SuperCDMS:2020aus,CRESST:2019jnq,Fuss:2022fxe}, with event rates decreasing over time after cooldown~\cite{Fuss:2022fxe,EDELWEISS:2016nzl,PyleExess,StrandhagenExess}. This behavior is likely due to the relaxation of thermal stress arising from the different contraction of detector materials and support structures during cooling~\cite{anthony2024stress}, which can release phonon bursts in the detector. These bursts typically deposit energies above the eV scale, with rates decaying exponentially over a timescale of 6--10 days~\cite{anthony2024stress,Yelton:2025wsy}. Such backgrounds can be mitigated by waiting sufficiently long after cooldown for thermal equilibrium to be restored, fabricating and transporting detectors under cryogenic conditions to minimize stress formation, or employing specialized low-stress support designs~\cite{anthony2024stress,Yelton:2025wsy,CRESST:2019jnq,Astrom:2005zk}.
    
\end{itemize}

\section{Random-error-induced dark counts}
We now deal with the dark counts arising from imperfect readout. Consecutive relaxation or readout errors can, in principle, mimic a parity flip. We show here that such events contribute negligibly to the dark count rate after proper signal processing and threshold choice, assuming 95\% overall fidelity.

\begin{figure}[!htp]
    \centering
    \includegraphics[width=1.0\columnwidth]{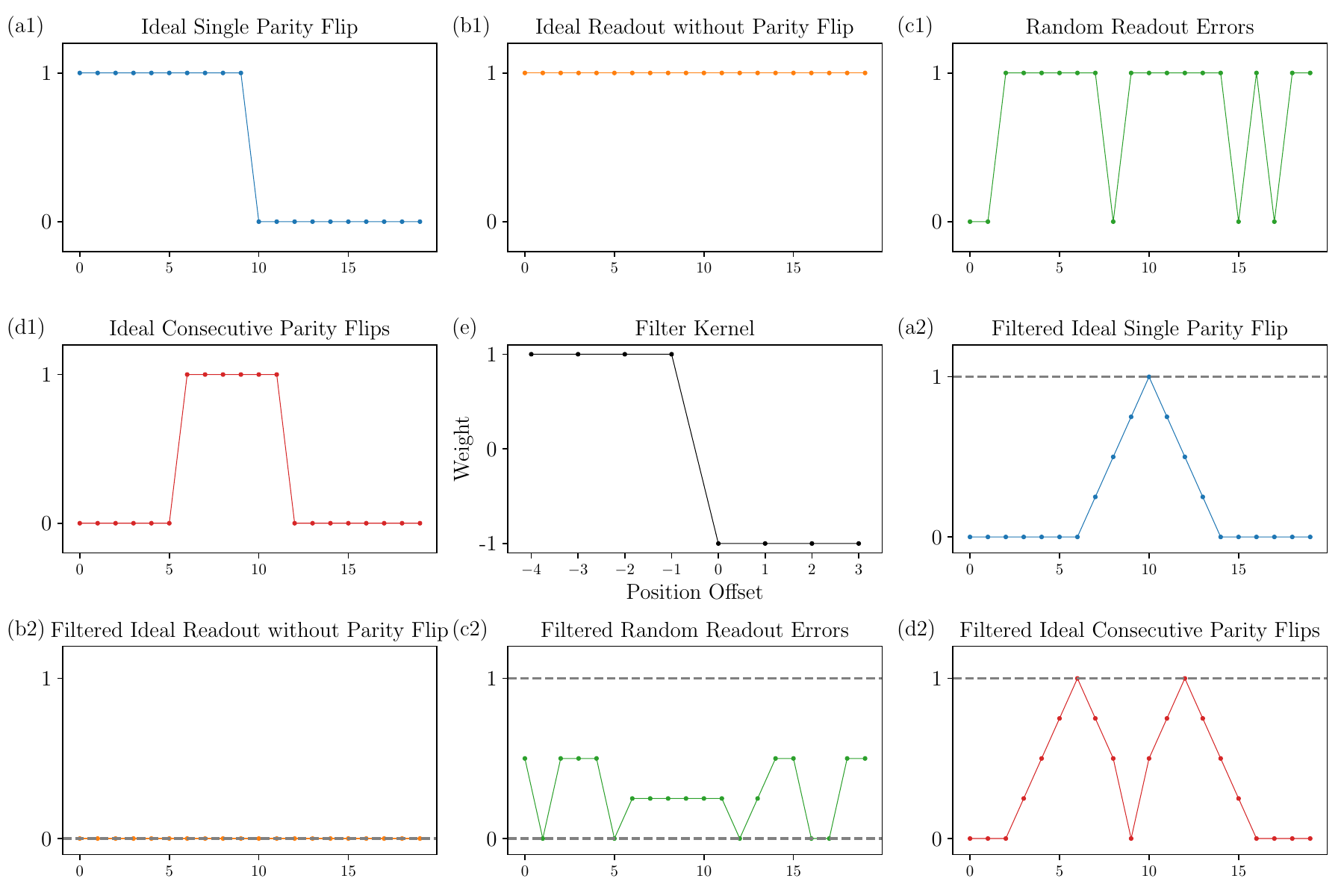}
    \caption{Schematic illustrations of parity flip signal processing.
\textbf{(a1)--(d1):} Four representative raw signal patterns: 
(a1) ideal single parity flip (step function), 
(b1) ideal readout without parity flips (constant signal), 
(c1) random readout errors (random 0's superimposed on constant 1's), 
and (d1) ideal readout with two consecutive parity flips.
\textbf{(e):} The $L=4$ filter kernel $h = [1,1,1,1,-1,-1,-1,-1]$.
\textbf{(a2)--(d2):} Normalized filter outputs, see text for details. The filter effectively detects real parity flips by producing peaks at transition regions, 
while suppressing constant signals (b2).
}
    \label{fig:filter}
\end{figure}

In Fig.~\ref{fig:filter} we show the different behaviors of real signals and readout errors. While a real parity flip exhibits different parity states before and after the flip, a readout error has the same parity state before and after the error occurs. The different behaviors can be distinguished by applying a proper filter.  We adopt a filter kernel that consists of $L$ $1$'s followed by $L$ $-1$'s, i.e., 
$h=[\underbrace{1, \dots, 1}_{L}, \underbrace{-1, \dots, -1}_{L}]$ of length $2L$. This kernel is then correlated with the raw data $s[i]=0\,\text{or}\,1$ corresponding to different parity states. We then take the absolute value of the outcome and normalize it by dividing it by $L$:
\begin{equation}
    s'[k]=\frac{1}{L}\left|\sum_{i=0}^{2L-1} s[k+i-L]h[i]\right|.
    \label{eq:matched_filtering}
\end{equation}
For a perfect step signal from a single parity flip, the procedure above yields a peak value of $1$ for the filtered output $s'$; see Fig.~\ref{fig:filter}. For a signal that is constantly $1$ or $0$, the result is $0$, and for signals containing random errors, the result lies between $0$ and $1$.

We determine the threshold for signal/error selection through a large amount of simulations. In the absence of any true parity flips, i.e., for raw data consisting solely of 5\% random noise, we perform a peak search on the filtered output $s'$ for $N=10^{11}$ samples ($k$ ranges from $0$ to $N-1$ in Eq.~\ref{eq:matched_filtering}) with $L=12$ filter. Assuming a measurement interval of $1\,\mu\text{s}$, the total observation time of the simulation run is $10^5\,\text{s}$. The histogram of the resulting peak values is shown in Fig.~\ref{fig:hist}. We calculate probability of two types of events:
\begin{figure}[!htp]
    \centering
    \includegraphics[width=0.6\columnwidth]{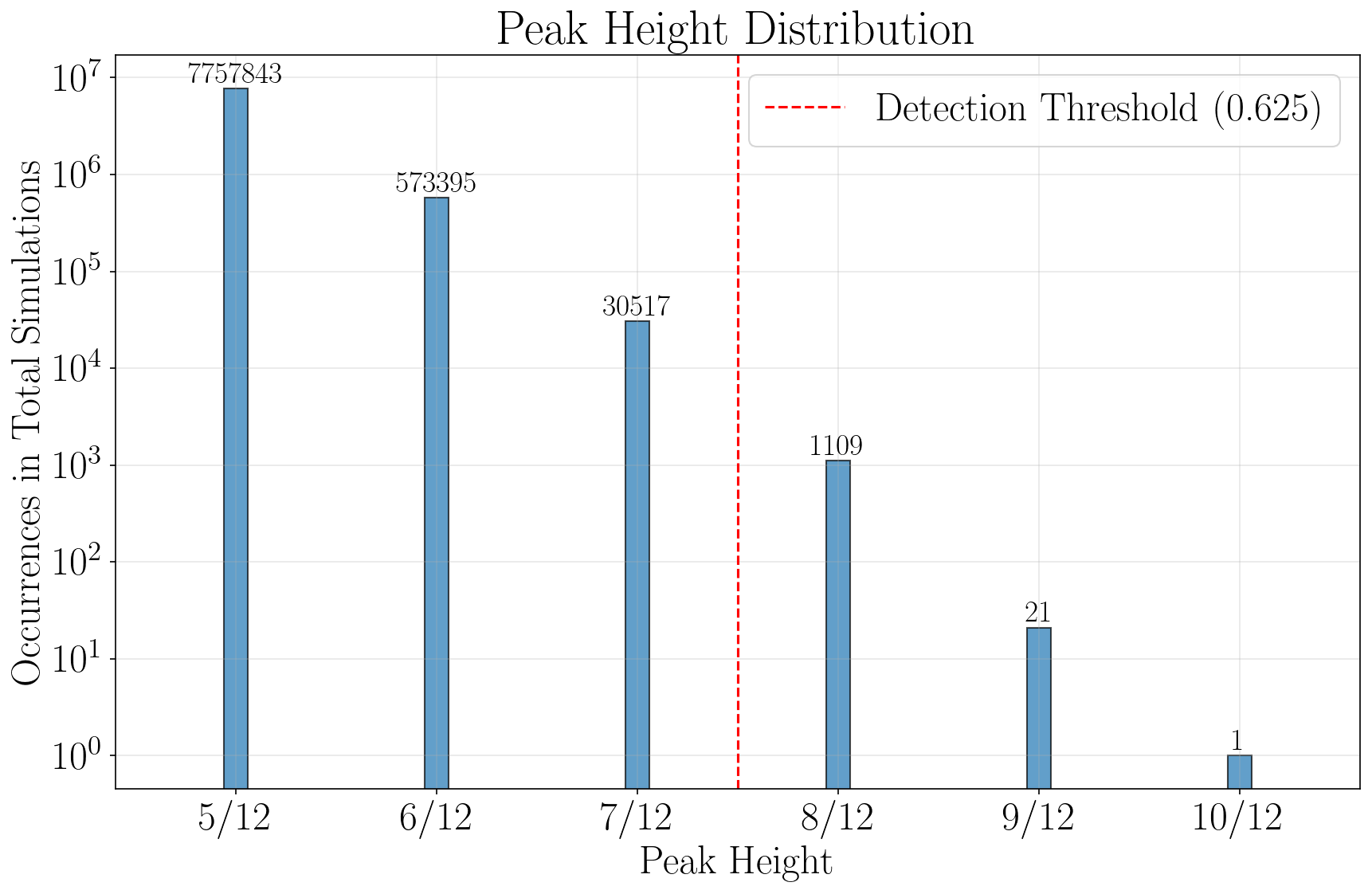}
    \caption{Histogram of peak values obtained from filtered output of $N=10^{11}$ samples simulated assuming 5\% random noise without real signal. The horizontal axis shows the normalized peak amplitude from $5/12$ to $10/12$. No peak height values above $10/12$ are found in the simulation.}
    \label{fig:hist}
\end{figure}

\begin{itemize}
    \item {\it False positive.} This is the probability for the readout noise to mimic a parity flip signal. We show the occurrences of different peak values in all $N=10^{11}$ simulations in Fig.~\ref{fig:hist} with readout noise. Due to the length of the filter and the filtering method, the peak values can only take discrete fractional numbers from 0 (0/12) to 1 (12/12). As expected, the number of occurrences decreases with increasing peak height, and we observe no event with peak height larger than 10/12. We therefore choose the selection threshold of $0.625$ and require the peak height to be larger than $0.625$ to be identified as a parity flip signal. This corresponds to a false-positive probability of $1.13\times 10^{-8}$, or considering the simulation time, a false-positive rate of 0.01 Hz. 
    For a conservative estimate, if the rate of false signals induced by such random errors is an order of magnitude lower than the background parity‑flip rate due to residual quasiparticle tunneling, then its contribution can be neglected. For a parity‑flip rate of $1\,\text{Hz}$ (corresponding to a quasiparticle density of about $10^{-3}\,\mu\text{m}^{-3}$, which is the benchmark case in the main text; see Fig.~2 in the main text), the false‑signal rate from random errors should be suppressed below $0.1\,\text{Hz}$. The false‑signal rate from our selection cut is two orders of magnitude lower than residual quasiparticle tunneling, and will not affect the results.
    \item {\it False negative.} This is the probability for a parity flip signal to be falsely identified as readout noise due to event selection. The probability that a true step signal from a parity flip yields a filtered peak amplitude below the detection threshold of $0.625$ can be estimated by a simple calculation of the binomial distribution. For a perfect step signal processed by the $L=12$ filter, the $2L = 24$ binary samples within the window consist of twelve $1$'s followed by twelve $0$'s (or vise versa). Assuming an independent random error rate of $5\%$ per sample, the number of errors in the window follows a binomial distribution $X\sim B(n=24, p=0.05)$. A filtered output below $0.625$ occurs only when five or more errors are present. The corresponding probability is $P(X \ge 5) \simeq 0.006$. Such low false-negative probability will not reduce the real event rate in a noticeable way.
\end{itemize}

One may also worry about the fact that the above selection process may not work when two parity flips are too close to each other in time. In fact, as long as the time interval between two successive parity flips exceeds the half filter length $L$, or $12\,\mu\text{s}$ in the case of $L=12$ and a measurement interval of $1\,\mu\text{s}$, the above procedure remains effective; see Fig.~\ref{fig:filter} (a2 and d2). This condition amounts to less than $8\times10^3$ expected tunneling counts for 96 qubits assuming a 1-ms quasiparticle decay lifetime, an upper limit that is much larger than the expected number of tunneling counts indicated in Fig. 2 in the main text for the dark matter energy deposition within our region of interest.


\end{document}